\documentclass[aps,prb,twocolumn,superscriptaddress,showpacs,reprint]{revtex4-1}
\usepackage{graphicx}
\usepackage{mathrsfs}
\usepackage{amsfonts}
\usepackage{amsmath}
\usepackage{color}
\usepackage{xcolor}
\usepackage[colorlinks=true,letterpaper=true, pdfstartview=FitV,urlcolor=blue,linkcolor=orange,citecolor=cyan]{hyperref}
\usepackage{subfigure}
\usepackage{bm}
\usepackage{verbatim}
\usepackage{color}
\newcommand{\bk}{\bm{k}}
\newcommand{\bE}{\bm{E}}
\newcommand{\partii}[2]{\dfrac{\partial #1}{\partial #2}}
\newcommand{\rk}{\rho}
\newcommand{\rko}{\rho_{0}}
\newcommand{\rklo}{\rho_{1}^{(0)}}
\newcommand{\rkle}{\rho_{1}^{(e)}} 
\newcommand{\rkl}{\rho_{1}}
\newcommand{\fkl}{f}
\newcommand{\fklo}{f^{(0)}}
\newcommand{\fkle}{f^{(e)}}
\newcommand{\drudeo}{\mathcal{D}^{(0)}}
\newcommand{\drudee}{\mathcal{D}^{(e)}}
\newcommand{\dic}{\bar{\mathcal{D}}}
\newcommand{\betaa}{\beta_{1}}
\newcommand{\betab}{\beta_{2}}
\newcommand{\mH}{\mathcal{H}}
\newcommand{\eps}[1]{\varepsilon_{#1}(\bm{k})}
\newcommand{\delk}{\Delta_{\bk}}
\newcommand{\delkp}{\Delta_{\bk'}}
\newcommand{\varf}{\varepsilon_F}
\newcommand{\ktf}{k_{\text{TF}}}
\newcommand{\kbig}{k_{F+}}
\newcommand{\ksmall}{k_{F-}}
\newcommand{\tdk}{\tilde{k}}
\newcommand{\tdkbig}{\tilde{k}_{F+}}
\newcommand{\tdksmall}{\tilde{k}_{F-}}
\newcommand{\ellk}{\mathbb{K}}
\newcommand{\elle}{\mathbb{E}}
\newcommand{\epa}{e(\bm{E}\cdot\hat{\bm{k}})}
\newcommand{\epb}{e(\bm{E}\times\hat{\bm{k}})_z}

\newcommand{\ek}{\epsilon_{\bk}}
\newcommand{\nn}{\hat{\bm{n}}}
\newcommand{\bsigma}{\bm{\sigma}}
\newcommand{\Paulia}{\bsigma\cdot\nn}
\newcommand{\Paulib}{(\bsigma\times\nn)_z}
\newcommand{\mA}{\mathcal{A}}
\newcommand{\fkk}{\phi_{k'k}}
\newcommand{\ket}[1]{|#1\rangle}
\newcommand{\bra}[1]{\langle#1|}
\newcommand{\se}{\Sigma^{(0)}}
\newcommand{\dse}{\delta\Sigma} 
\usepackage{siunitx}
\begin{document}
\title{Theory of interaction-induced renormalization of Drude weight 
and plasmon frequency in chiral multilayer graphene}
\author{Xiao Li}
\affiliation{Condensed Matter Theory Center and Joint Quantum Institute, University of Maryland, College Park, MD 20742, USA}
\affiliation{Department of Physics, The University of Texas at Austin, Austin, Texas 78712, USA}
\author{Wang-Kong Tse}
\affiliation{Department of Physics and Astronomy, Center for
Materials for Information Technology, The University of Alabama, Alabama 35487, USA}

\date{\today}

\begin{abstract}
We develop a theory for the optical conductivity of doped multilayer graphene including the effects of electron-electron interactions.  
Applying the quantum kinetic formalism,
we formulate a set of pseudospin Bloch equations that governs the dynamics of the nonequilibrium density matrix driven by an external \emph{a.c.} electric field under the influence of Coulomb interactions. These equations reveal a dynamical mechanism that couples the Drude and interband responses arising from the chirality of pseudospin textures  in multilayer graphene systems. We demonstrate that this results in an interaction-induced enhancement of the Drude weight and plasmon frequency strongly dependent on the pseudospin winding number.   
Using bilayer graphene as an example, we also study the influence of higher-energy bands and find that they contribute considerable renormalization effects not captured by a low-energy two-band description. We argue that this enhancement of Drude weight and plasmon frequency occurs generally in materials characterized by electronic chirality. 
\end{abstract}
\pacs{81.05.ue, 78.67.Wj, 71.10.-w, 78.67.Pt}
\maketitle

\section{Introduction}
Galilean invariance is generally broken in solids due to the presence of a lattice background. For typical semiconductor materials however, Galilean symmetry is preserved for low-energy states near the band edge where the only remaining effect of the underlying lattice is a renormalization of the electron mass from its bare value~\cite{Yu:2010Springer_Semiconductors}. In a  Galilean-invariant system, interaction effects do not affect electronic transport which is only carried by the center-of-mass motion of the electron liquid. The absence of interaction corrections to Drude weight in conventional two-dimensional electron gas (2DEG) has been demonstrated in several experiments~\cite{Hirjibehedin:2002is,Pellegrini:2006el}.

On the other hand, electronic transport in multilayer graphene systems is incompatible with Galilean invariance symmetry due to the chiral pseudospin texture of their low-energy states. Electronic states in the Brillouin zone are not only characterized by their respective crystal momenta, but also by their pseudospin orientations that originate from the underlying lattice structure. A Galilean boost in graphene systems will not only shift the momentum of the occupied quantum states but also  change  their average pseudospin orientations. As a result, electronic states in chiral multilayer graphene do not respect Galilean symmetry. Therefore, unlike conventional  2DEG, optical properties of graphene systems can be subject to renormalization effects from many-body interactions~\cite{James:2009PRB_TwobandModel,Polini:2011PRB_Drudeweight}. 

The above theoretical expectations receive reasonable support from the experiments but remain an open issue to date. 
In single-layer graphene, several measurements of the  Drude weight indeed observe a deviation from its free-carrier behavior, though the experimental {interpretations are not yet fully conclusive}. 
In two earlier optical spectroscopy  experiments~\cite{Horng:2011PRB_DrudeCond,Yan:2011ACS_DrudeCond}, the results suggest an up to $40\%$ suppression of the Drude weight. However, in a recent cyclotron-resonance absorption experiment~\cite{Orlita:2012NJP_Drude}, the measured Drude weight is reported to be in quantitative agreement with the prediction  in Ref.~\onlinecite{Polini:2011PRB_Drudeweight}. Optical Drude weight in bilayer graphene is less studied  experimentally. To date, most optical absorption measurements on  bilayer graphene focus on the higher-frequency absorption features in the spectrum such as interband absorption thresholds as well as the asymmetry between electron- and hole-doped regions~\cite{Abergel:2007PRB_OpticalInfrared,Nicol:2008PRB_BLGCond,Nilsson:2008PRB_ElePropertyBLG,Benfatto:2008PRB_OpticalSumRule,Wang:2008Science_GateTunableTransition,Malard:2007PRB_Raman,Zhang:2008PRB_BLGconductivity,Kuzmenko:2009PRB_InfraredSpect,Li:2009PRL_BandAsymmetry,Zhang:2009Nature_TunableGap,Mak:2009PRL_BandGapOpening,Lui:2011NatPhys_BandgapTuning}. Additional studies are imperative to better understand the intraband absorption processes represented by the optical Drude weight in bilayer and multilayer graphene. 

In this paper, we present a quantum kinetic theory for the renormalization of the Drude weight and plasmon frequency in multilayer graphene. The current work generalizes the theory developed in Ref.~\onlinecite{James:2009PRB_TwobandModel} for the case of bilayer graphene. 
The quantum kinetic approach~\cite{Rammer:1986RMP_QKE} captures important quantum coherence effects among energy bands beyond the semiclassical Boltzmann theory. 
We first build our theory on the low-energy two-band description of multilayer graphene and study the effects of chirality on the interaction-induced renormalization. Using the full four-band Hamiltonian, we then focus on bilayer graphene as an example to illustrate the effects of higher energy bands ignored in the two-band model. In these two calculations, we obtain a set of generalized optical Bloch equations that govern the dynamical frequency dependence of the nonequilibrium density matrix under the influence of an optical field and electron-electron interactions. We obtain leading-order solutions to these equations and demonstrate that the Drude weight and plasmon frequency are enhanced, with substantial corrections from higher-band contributions that are ignored in the two-band calculations.   

We organize the rest of our paper as follows. We first develop the formalism of our kinetic theory for multilayer graphene using the two-band model in Section~\ref{Sec:Model} and~\ref{Sec:SLGDIC}. Then in Section~\ref{Sec:4bandModel} we lay out the necessary ingredients for a more elaborate theory for bilayer graphene using the four-band description. In Section~\ref{Sec:Drudeweight} we proceed to obtain the leading-order solution to the theory and obtain the optical Drude weight of bilayer graphene. 
In Section~\ref{Sec:Discussions} we compare and discuss the results obtained using the two-band and four-band models of bilayer graphene, as well as the renormalization of the plasmon frequency. 
Finally, Section~\ref{Sec:Conclusion} summarizes our main results. 

\section{Quantum kinetic formalism \label{Sec:Model}}

We make use of a quantum kinetic equation~\cite{Rammer:1986RMP_QKE} to study the influence of electron-electron interactions on the optical conductivity. 
Such an approach is well established in connection with studies of the carrier and exciton kinetics in conventional semiconductors under optical excitation~\cite{haug2008quantum}. 
While fully equivalent to the Bethe-Salpeter equation, the advantage of the present approach lies in the gauge invariance structure of the kinetic equation, in which electron self-energy effects as well as excitonic effects are built in consistently in a conserving approximation. The density matrix $\rk$ is the central quantity in our theory. In the presence of an \textit{a.c.} electric field $\bE$, the dynamics of the density matrix $\rk=\rk(\bk)$ is governed by the following quantum kinetic equation~\cite{James:2009PRB_TwobandModel}
\begin{align}
-i\omega\rk+e\bE\cdot\partii{\rk}{\bk}+i[\mH,\rk]=0, \label{Eq:QKE-general}
\end{align}
where $\omega$ is the frequency of the \textit{a.c.} field and $[\hat{A},\hat{B}]\equiv \hat{A}\hat{B}-\hat{B}\hat{A}$ denotes the commutator between operators $\hat{A}$ and $\hat{B}$. The system Hamiltonian $\mH$ generally comprises a noninteracting part $\mH_0$ and a self-energy correction due to many-body interactions. In linear response, the density matrix is given by $\rk=\rko+\rkl$, where $\rko$ is its equilibrium value and $\rkl$ the first-order correction due to the external electric field. Keeping only terms up to the first order in electric fields, we write Eq.~\eqref{Eq:QKE-general} as 
\begin{align}
 -i\omega\rkl + e\bE\cdot\partii{\rko}{\bk}+i[\mH,\rko+\rkl]=0. \label{Eq:QKE}
\end{align}
Our focus is on obtaining the Drude weight from the optical conductivity. Because the Drude weight is obtained from  the residue of the $\omega=0$ pole in the real part of the optical conductivity in the absence of disorder, we can limit our discussions to the clean limit $\omega\tau\gg1$, where collision terms in the kinetic equation can be  ignored. The optical conductivity $\sigma(\omega)$ can then be obtained from the average total current $J = \sigma(\omega)\bE$, which is the quantum mechanical average of the current operator $j$  
\begin{align}
  J &= g_sg_v e\sum_{\bk}\;\text{tr}(\rkl j).\label{Eq:totalcurrents}
\end{align}
In the above equation, $g_s,g_v = 2$ arise from the spin and valley degeneracies respectively in multilayer graphene systems and `tr' denotes trace over the pseudospin degrees of freedom. In the following, we will solve for the nonequilibrium density matrix $\rkl$ from Eq.~\eqref{Eq:QKE} both in the absence and presence of electron-electron interaction. To clearly delineate these two limits, we separate $\rkl$  into two parts  $\rkl=\rklo+\rkle$, with the first term $\rklo$ being the noninteracting result and the second term $\rkle$ containing corrections from interaction effects. 


\section{Chiral multi-layer graphene \label{Sec:SLGDIC}}

In this section, we generalize the method {used for obtaining the Drude weight renormalization} developed in Ref.~\onlinecite{James:2009PRB_TwobandModel} from the case of bilayer to multilayer graphene. By focusing our interest on the two lowest energy bands around the charge neutrality point, one can write~\cite{Min:2008PRB_PseudospinMag,McCann:2011PRB_MultilayerGraphene,Li2014} the effective Hamiltonian for an $l$-layer ABC-stacked multilayer graphene system as $\mH_0=\ek\nn\cdot\bsigma$, where $\nn=(\cos l\phi_k,\sin l\phi_k)$ is the pseudospin vector  responsible for the chirality of the band structure, $\bsigma$ is a vector comprising the set of Pauli matrices acting on the pseudospin degrees of freedom, $\ek\equiv \mA_l k^l$ is the band energy dispersion, with $\mA_l=(\hbar v_0)^l/\gamma_1^{l-1}$. 
In this low-energy description, the pseudospin degrees of freedom correspond to the outermost top and bottom layers in multilayer graphene (including bilayer graphene)  
and the two sublattice sites for single-layer graphene. As the electronic wave vector undergoes one full rotation around the Dirac point, the pseudospin vector also undergoes $l$ number of rotations. In other words, the pseudospin winding number is equal to the number of layers~\cite{Park:2011ps}. 
We note that this two-band model is valid within a limited energy range; in particular for bilayer and multilayer graphene it does not capture either the higher-energy bands or the low-energy remote hopping processes that can lead to trigonal warping effects~\cite{Varlet2014:BLG}. 


\subsection{Pseudospin Bloch Equation}

In equilibrium, the density matrix in the band basis is diagonal with the elements $n_F(\xi_{k\lambda})$, where $n_F(x)$ is the Fermi-Dirac distribution function,  $\xi_{k\lambda}=\lambda\ek-\varf$ is the quasiparticle energy measured from the Fermi energy $\varf$, and $\lambda=+(-)$ labels the conduction (valence) band. For clarity we denote $n_F(\xi_{k\lambda})$ simply by $n_\lambda(k)$ in the following. The equilibrium density matrix $\rko$ can be obtained by transforming the above diagonal matrix from the band basis to the pseudospin basis yielding 
\begin{align}
	\rko = \dfrac{1}{2}\sum_{\lambda=\pm} n_\lambda(k)(1-\lambda\bsigma\cdot\nn). 
\end{align}

To obtain the nonequilibrium density matrix $\rk$, we first express it in the complete basis of a set of transformed Pauli matrices (see Appendix~\ref{Appendix:GammaMatrices})  
%
\begin{align}
	\rkl &= i\Paulib P(\bk)+Q(\bk)\sigma_z+R(\bk)\Paulia+S(\bk). \label{Eq:TwobandRk}
\end{align}
Such a decomposition carries a clear physical meaning: the $\bm{1}$, $\Paulia$, $\Paulib$, and $\sigma_z$ components describe the total density change,  interband polarization, interlayer coherence, and interlayer polarization respectively.  

We insert the above ansatz for $\rkl$ into the quantum kinetic equation in Eq.~\eqref{Eq:QKE} and obtain the following equations for the functions $P$, $Q$, $R$, and $S$. 
In particular, $R(\bk)$ and $S(\bk)$ have the following closed form solutions
\begin{align}
	S(\bk)&=-\dfrac{i\epa}{2\omega}\left[n'_+(k)+n'_-(k)\right],\notag\\
	R(\bk)&= \dfrac{i\epa}{2\omega}\left[n'_+(k)-n'_-(k)\right], \label{Eq:Two-band-Bk}
\end{align} 
while $P(\bk)$ and $Q(\bk)$ satisfy the following coupled integral equations, 
\begin{align}
	\omega P(\bk)+\delta_k Q(\bk)&=(n_{+}-n_{-})\left[\dse_{+-}+e\bm{E}\cdot\mathcal{A}_{+-}\right],\notag\\
	\delta_k Q(\bk)+\omega P(\bk)&=-(n_{+}-n_{-})\dse_{-+},\label{Eq:two-bandEquations} 
\end{align}
where $\delta_k=2\ek+\Sigma_{+}^{(0)}(\bk)-\Sigma_{-}^{(0)}(\bk)$ is the interband excitation energy, and 
\begin{align}
	\Sigma_{\lambda}^{(0)}(\bk)=-\sum_{\lambda'=\pm,\bk'}V_{\bk\bk'}n_{\lambda'}(k')\left[1+\lambda\lambda'\cos (l\fkk)\right]/2\notag
\end{align}
is the equilibrium self-energy for band $\lambda=\pm$. The electric dipole term consists of a coupling between the electric field and a gauge potential $\mathcal{A}_{+-}$. 
Here $\mathcal{A}_{\lambda\lambda'}$ is the \emph{non-Abelian Berry connection}~\cite{DiXiao:2009RMP_BerryPhase}, 
\begin{align}
	\mathcal{A}_{\lambda\lambda'}(\bk) = i\bra{u_{\lambda}(\bk)}\frac{\partial}{\partial \bk}\ket{u_{\lambda'}(\bk)},\label{Eq:BerryConnection}
\end{align}
with $u_{\lambda}(\bk)$ denoting the wave function for the band $\lambda$. $\mathcal{A}_{+-}$ is therefore the off-diagonal matrix element of $\mathcal{A}(\bk)$ between the conduction and valence band states, and for multilayer graphene we have $\mathcal{A}_{+-} =(l/2k)\hat{\phi}$ in Eq.~\eqref{Eq:two-bandEquations}. 
Finally, the right-hand side of Eq.~\eqref{Eq:two-bandEquations} arises from changes in the self-energy from the nonequilibrium density matrix 
\begin{align}
	\dse_{+-}(\bk) &=\sum_{\bk'}V_{\bk\bk'} Q(\bk'), \label{Eq:DrudeContributionToCk}\\
	\dse_{-+}(\bk) &=\sum_{\bk'}V_{\bk\bk'}[\cos l\fkk P(\bk')-i\sin l\fkk R(\bk')].\notag
\end{align}
%

Eqs.~\eqref{Eq:Two-band-Bk}-\eqref{Eq:two-bandEquations} comprise a set of pseudospin Bloch equations, reminescent of the optical Bloch equations commonly used in two-level atoms \cite{allen2012optical} and conventional two-band  semiconductors~\cite{haug2008quantum,boyd2013nonlinear}. Importantly,  Eqs.~\eqref{Eq:Two-band-Bk}-\eqref{Eq:two-bandEquations} are also different from the conventional optical Bloch equations in the following way.  
First we note that the solutions of $R(\bk)$ and $S(\bk)$ in Eq.~\eqref{Eq:Two-band-Bk} describe the Drude responses of the total density and interband polarization. Eq.~\eqref{Eq:two-bandEquations} determines the interband response from the coupled dynamics of the coherence and polarization in the layer degrees of freedom. An important observation is that the interband response is coupled to the Drude response through the nonequilibrium self-energy $\dse_{-+}(\bk)$ in Eq.~\eqref{Eq:DrudeContributionToCk} due to its dependence on $R(\bk)$. This Drude-interband coupling is the central piece of physics that gives rise to the renormalization effects on the optical conductivity and plasmon frequency we discuss in this paper. It arises from the exclusive $\hat{\phi}$ dependence in the Berry connection $\mathcal{A}_{+-}$ in Eq.~\eqref{Eq:two-bandEquations} that reflects the chirality of the graphene bandstructure.  


Solutions of Eq.~\eqref{Eq:two-bandEquations} yield the interaction corrections to the density matrix $\rkl$. 
To obtain the optical conductivity, we need to compute the current induced by an applied \emph{a.c.} electric field. 
We decompose the current density operator in the following way
\begin{align}
 \bm{j}_{\bk} = \partii{\mH_0}{\bk}=\partii{\mH_0}{k}\hat{k}+\dfrac{1}{k}\partii{\mH_0}{\phi}\hat{\phi}\equiv j_k\hat{k}+j_{\phi}\hat{\phi},
\end{align}
where $\hat{k}$ and $\hat{\phi}$ are the unit vector for the radial and azimuthal direction, respectively. 
The current density along the $x$-direction is then $ j_x= j_k\cos\phi - j_\phi \sin\phi$. 
As a result, in the linear response regime, the total current induced by an electric field in the $x$ direction reads
\begin{align} 
  J_x &= 4e\int d\bk\;\text{Tr}(\rkl j_x) \equiv J_1 - J_2,  \label{Eq:totalcurrents}
\end{align}
with $J_1$ and $J_2$ defined as
\allowdisplaybreaks[2]
\begin{align*}
   J_1= 4e\int d\bk\;\text{Tr}(\rkl j_k)\cos\phi, \;
  J_2 = 4e\int d\bk\;\text{Tr}(\rkl j_\phi)\sin\phi.
\end{align*}
For the two-band model we find that the current operator $j_x = \partial \mH_0/\partial k_x$ is evaluated as $j_x = l\mA_lk^{l-1}\{\sigma_x\cos[(l-1)\phi_k]+\sigma_y\sin[(l-1)\phi_k]\}$. As a result, the total current of the system is given by 
\begin{align}
	J_x = \dfrac{2e}{\pi^2}l\mA_l \int_0^{\infty} dk k^l\int_0^{2\pi} d\phi_k \left[R(\bk)\cos\phi_k+iP(\bk)\sin\phi_k\right], \label{Eq:Two-band-current}
\end{align}
from which we can find the interaction corrections to the conductivity.

\subsection{Leading-order interaction-induced Drude weight renormalization}

We now use the above formalism to obtain the leading-order interaction-induced Drude weight renormalization $\dic$ in multilayer graphene. The integral equations~\eqref{Eq:TwobandRk}-\eqref{Eq:two-bandEquations} can be solved numerically to obtain the nonequilibrium density matrix $\rkl$ to all orders of interaction potential within our theory. 
To maintain analytic tractability, {however}, in this work we will {only} solve these couped integral equations perturbatively up to first order and obtain the corresponding  interaction corrections to the Drude weight. 
{In addition}, as we are concerned only  with the Drude weight, it is sufficient to evaluate terms with an $\omega^{-1}$ dependence in Eq.~\eqref{Eq:Two-band-current}. 

First, the noninteracting contribution to the Drude weight only comes from $R(\bk)$ in Eq.~\eqref{Eq:Two-band-Bk}, yielding
\begin{align}
  \drudeo(\varf) = \dfrac{e^2}{\pi}l\mA_l k_F^l=\dfrac{e^2}{\pi}l\varf,
\end{align}
where $k_F$ is the Fermi wave vector. 
The interaction contributions to the Drude weight are contained in the $P(\bk)$ term from Eq.~\eqref{Eq:Two-band-current}, originating from the nonequilibrium self-energy $\dse_{-+}(\bk)$ due to Drude-interband coupling. To leading order in the interaction potential, we find that the part of $P(k)$ having a $\omega^{-1}$ dependence (denoted by a subscript `$\mathrm{Drude}$' below) is given by 
\begin{align}
	P_{\mathrm{Drude}}^{(e)}&= \dfrac{\epb(n_{+}-n_{-})}{4\ek}\times \notag\\
	&\sum_{\bk'}V_{\bk\bk'}\sin\fkk\sin l\fkk(n_{-}'-n_{+}').
\end{align}
The leading-order interaction correction to the Drude weight then follows from substituting the above into Eq.~\eqref{Eq:Two-band-current}. 

To illustrate the behavior of the Drude weight correction in the presence of screening effects, we assume static screening for the Coulomb potential 
\begin{align}
 V_{\bk\bk'}= \dfrac{2\pi e^2}{\kappa(|\bk-\bk'|+\eta\ktf)}, \label{Eq:CoulombPotential}
\end{align}
where $\kappa$ is the effective dielectric constant of the environment in which the multilayer graphene sheet is embedded, $\ktf$ is the Thomas-Fermi screening wave vector,  
\begin{align}
	\ktf=\dfrac{g_sg_ve^2}{l\kappa}\mA_l^{-1}k_F^{2-l}, 
\end{align}
and $\eta$ is a control parameter that can be adjusted to represent the weaker screening at high frequencies. 
In the following we first consider two limits where analytical expressions for the Drude weight correction can be obtained. We first study the long-range interaction limit corresponding to negligible screening by  ignoring the $\ktf$ in Eq.~\eqref{Eq:CoulombPotential}. In the opposite limit when the interaction is heavily screened, the Thomas-Fermi screening length $\ktf$ will be much larger than typical values of $|\bk-\bk'|$. Thus, we use a constant $V_0$ interaction to represent $ V_{\bk\bk'}$. Finally we evaluate the Drude weight correction numerically in the full Thomas-Fermi approximation [Eq.~\eqref{Eq:CoulombPotential}] and compare the results from the three cases. 




\subsubsection{Long-range interaction limit}
In the limit of long-range Coulomb potential, the expression for $P_{\mathrm{Drude}}^{(e)}(\bk)$ reads
\begin{align}
	P_{\mathrm{Drude}}^{(e)}(\bk)=\dfrac{e^3(\bE\times\hat{\bk})_zk_F(n_{+}-n_{-})}{16\pi\omega\ek\kappa}\notag\\\times
	[\Phi_{l-1}(k,k_F)-\Phi_{l+1}(k,k_F)],\notag
\end{align}
from which we obtain the following correction to the Drude weight from Eq.~\eqref{Eq:Two-band-current} 
\begin{align}
	\drudee= \dfrac{e^4l k_F}{8\pi^2\kappa}
	\int_{1}^{k_c/k_F}dx [\Phi_{l-1}(x,1)-\Phi_{l+1}(x,1)], 
\end{align}
where the function $\Phi_{l}$ is defined in Appendix~\ref{Appendix:AuxFunctions}. In addition, we have defined the dimensionless variable $x=k/k_F$, and $k_c$ is the momentum cutoff for which the two-band description for the low-energy multilayer graphene model remains valid, which is dependent on the number of layers. 
For Fermi energies with $k_F \ll k_c$ where the two-band description holds to a good approximation, the upper limit of the integral above becomes large and the value of the integral becomes independent of $k_F$. 
Therefore, unlike the noninteracting Drude weight, the power-law dependence on $k_F$ of the leading-order $\drudee$ is independent of the number of layers $l$. We define the Drude weight renormalization factor $\dic$ by $\dic-1= {\drudee}/{\drudeo}$, and find that in the long-range limit 
\begin{align}
 	\dic-1 
 	=&\dfrac{\alpha^{\ast}}{8\pi(\hbar v_0k_F/\gamma_1)^{l-1}}\notag\\&\times \int_{1}^{k_c/k_F}dx  [\Phi_{l-1}(x,1)-\Phi_{l+1}(x,1)],
\end{align}
where $\alpha^{\ast} = e^2/\kappa \hbar v$ is the effective fine structure constant in graphene, and $\kappa$ is the dielectric constant from the environment. 
Among graphene systems, we note that single-layer graphene ($l=1$) is special because the Drude weight renormalization factor is independent of electron density in this long-range interaction limit,
\begin{align}
	\dic-1\big|_\text{SLG}=\dfrac{\alpha^{\ast}}{8\pi}
	\int_{1}^{k_c/k_F}dx [\Phi_{0}(x,1)-\Phi_{2}(x,1)]. \label{Eq:SLG_lr_DIC}
\end{align}
This agrees with results obtained from the diagrammatic formalism up to the same leading order~\cite{Polini:2011PRB_Drudeweight}. 
For bilayer graphene $(l=2)$, we have 
\begin{align}
	\dic-1\big|_\text{BLG}=\dfrac{\alpha^{\ast}\gamma_1}{8\pi\hbar v k_F}
		\int_{1}^{k_c/k_F}dx [\Phi_{1}(x,1)-\Phi_{3}(x,1)],  \label{Eq:BLG_lr_DIC}
\end{align}
which agrees with the result previously obtained in Ref.~\onlinecite{James:2009PRB_TwobandModel}. 

\subsubsection{Short-range interaction limit}
We now turn to the limit of short-range interaction where electron-electron interaction is assumed to be a constant $V_0$. The expression for $P_{\mathrm{Drude}}^{(e)}(\bk)$ in this limit is
\begin{align}
	P_{\mathrm{Drude}}^{(e)} &=\dfrac{e(\bE\times\hat{\bk})_z(n_+-n_-)k_F V_0}{32\pi^2\omega\ek}\notag\\
	&\times\int_0^{2\pi}d\phi\left\{\cos[(l-1)\phi_{\bk}]-\cos[(l+1)\phi_{\bk}]\right\}. \notag
\end{align}
Interestingly, we note that when the number of layers $l>1$, the above expression vanishes due to azimuthal symmetry. This finding generalizes our previous result~\cite{James:2009PRB_TwobandModel} for bilayer graphene to $l > 2$ multilayer graphene. Therefore $\dic$ vanishes in the short-range interaction limit for pseudospin winding number $l \geq 2$. 
Single-layer graphene ($l=1$) is special as only it has a nonzero  leading-order $\dic$ in the short-range limit. If we let the effective interaction strength to be $V_0 = 2\pi e^2/\kappa\eta\ktf$, the Drude weight correction is then
\begin{align}
	\drudee\big|_\text{SLG} =\dfrac{e^4k_F}{4\pi\kappa\eta\ktf}(k_c-k_F)
	=\dfrac{e^2\hbar v}{16\pi\eta}(k_c-k_F),
\end{align}
and the corresponding $\dic$ is 
\begin{align}
	\dic-1\big|_\text{SLG}=\dfrac{\eta}{16}\left(\dfrac{k_c}{k_F}-1\right), \label{Eq:SLG_sr_DIC}
\end{align}
in agreement with the result obtained in Ref.~\onlinecite{Polini:2011PRB_Drudeweight}. 

\subsubsection{Thomas-Fermi Screening}
\begin{figure}[!]
\centering
\includegraphics[scale=0.3]{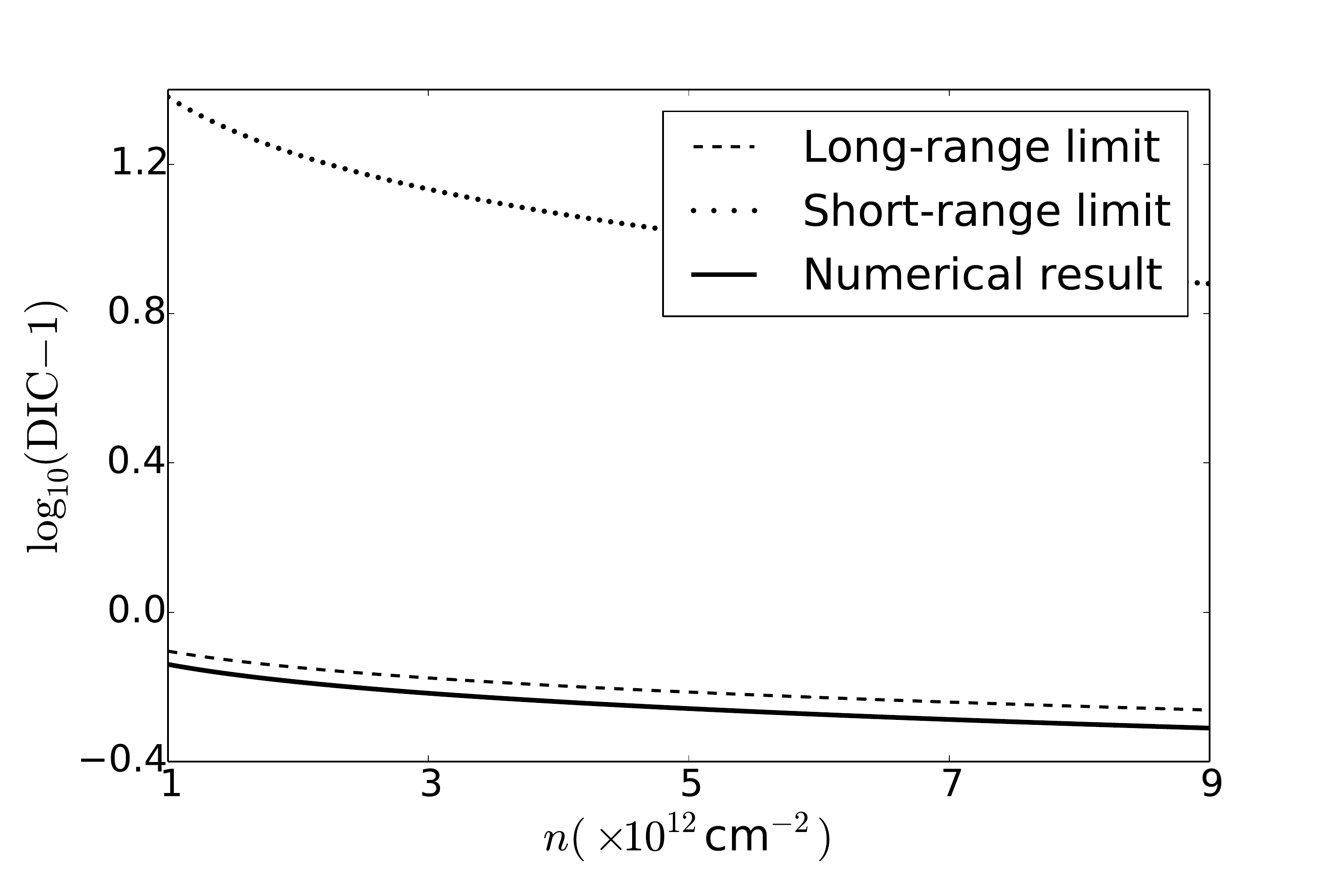}
\caption{Comparison of DIC in monolayer graphene in the long-range limit and short-range limit. We also include the numerical evaluation with the whole screened potential for comparison. Here we again use $\eta = 0.1$, and a dielectric constant of $\kappa =2.5$. \label{Fig:SLGDIC}}
\end{figure} 

We now evaluate the Drude weight renormalization numerically for finite static screening. The expression for the Drude weight correction for finite $\ktf$ is given by 
\begin{align}
	\drudee =\dfrac{e^4l}{4\pi^2\kappa}\int_{k_F}^{k_c}dk I_l(k),
\end{align}
with 
\begin{align}
	I_l(k)=\int_0^{2\pi}d\phi\dfrac{k_F}{|\bk-\bk_F|+\eta\ktf}\sin\phi\sin l\phi.
\end{align}
from which we obtain the Drude weight renormalization factor as 
\begin{align}
	\dic-1 =\dfrac{\alpha^{\ast}}{4\pi(\hbar v_0k_F/\gamma_1) ^{l-1}}\int_{1}^{k_c/k_F}dx I_l(x), \label{Eq:TwobandDIC}
\end{align}
In Fig.~\ref{Fig:SLGDIC}, we show the numerical result from Eq.~\eqref{Eq:TwobandDIC} and the analytical results in the long-range [Eq.~\eqref{Eq:SLG_lr_DIC}] and short-range limits [Eq.~\eqref{Eq:SLG_sr_DIC}]. We note that the short-range limit result drastically overestimates the Drude weight renormalization as compared to the Thomas-Fermi screening result, which is better approximated by the long-range limit. 

\begin{figure}[!]
\centering
\includegraphics[scale=0.3]{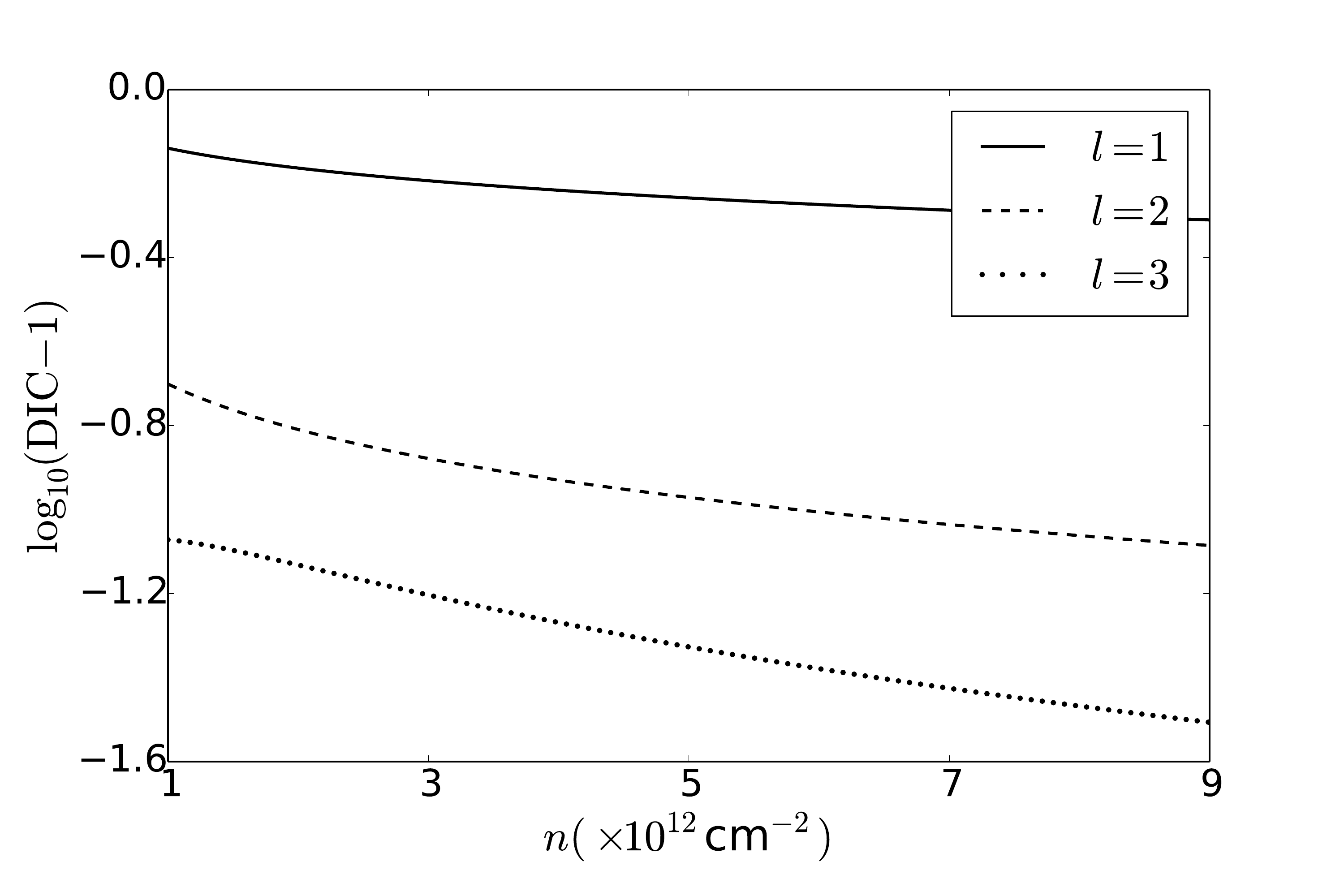}
\caption{Interaction-induced Drude weight renormalization in few-layer graphene [see the definition in Eq.~\eqref{Eq:TwobandDIC}]. Here we use $\eta = 0.1$, and the dielectric constant $\kappa =2.5$. \label{Fig:DICTwoband}}
\end{figure} 

Our theory further predicts that Drude weight renormalization effects become smaller with increasing number of layers, as shown in Fig.~\ref{Fig:DICTwoband}. Also, an increase in electron density will tend to weaken the Drude weight renormalization. 


\section{Generalization to four bands \label{Sec:4bandModel}}

\begin{figure}[!]
\includegraphics[scale=1]{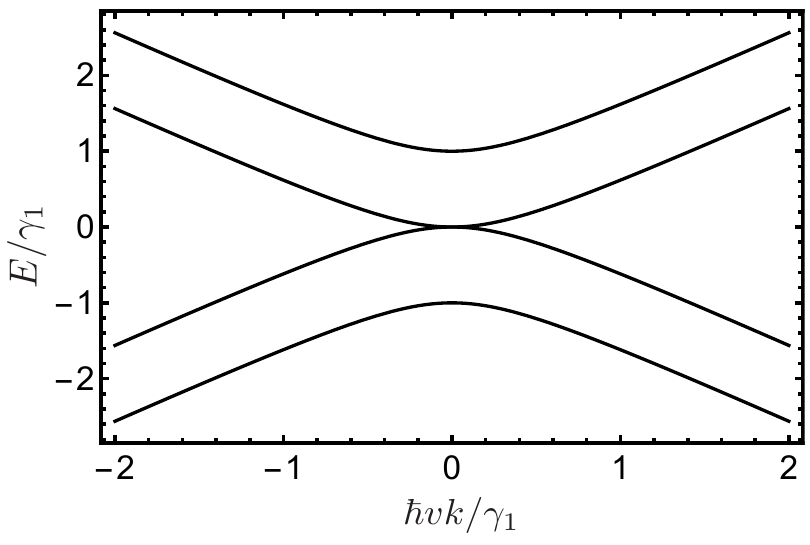}
\caption{\label{Fig:BLGBands} Bandstructure of bilayer graphene. In this figure $\gamma_1 = \SI{0.4}{eV}$ is the interlayer hopping energy, which is also equal to the energy  difference between the two conduction bands as well as that between the two valence bands. }
\end{figure}

In this section we generalize our kinetic equation formalism to more than two bands, using the $4\times4$ bilayer graphene model as a prototypical example. This serves to extend the validity of our theory to a wider frequency range encompassing higher frequency optical excitations, and to include the interband coherence effects between the two conduction bands as well as the two valance bands. Our starting point is the four-band continuum description of Bernal-stacked bilayer graphene, in which we only include the in-plane hopping energy and the nearest-neighbor interlayer coupling. The resulting Hamiltonian is given by~\cite{CastroNeto:2009RMP_Graphene}
\begin{align}
	\mathcal{H}_0=
	\begin{pmatrix}
		0 & v\hbar k e^{-i\phi} & -\gamma_1 & 0\\
		v\hbar k e^{i\phi} & 0 & 0 & 0\\
		-\gamma_1 & 0 & 0 & v\hbar k e^{-i\phi}\\
		0 & 0 & v\hbar k e^{i\phi} & 0
	\end{pmatrix},\label{Eq:BilayerH0}
\end{align}
where $v=1.0\times 10^6$\,m/s is the Fermi velocity of the Dirac fermions in single-layer graphene, $\phi=\tan^{-1}(k_y/k_x)$, and $\gamma_1=0.4$\, eV is the interlayer hopping energy. We will set $\hbar=1$ and $v=1$ hereafter and only restore them in our final results. The four bands derived from the above Hamiltonian are 
\begin{align}
	\eps{1}=\dfrac{1}{2}\left(\sqrt{4k^2+\gamma_1^2}+ \gamma_1 \right)=-\eps{4},\notag\\
	\eps{2}=\dfrac{1}{2}\left(\sqrt{4k^2+\gamma_1^2}- \gamma_1 \right)=-\eps{3}, \label{Eq:BLband}
\end{align}
which is shown in Fig.~\ref{Fig:BLGBands}, and the corresponding wavefunctions will be denoted as $u_{i}(\bk)$. 
For convenience, we will adopt the notation $\delk\equiv\sqrt{4v^2k^2+\gamma_1^2}$ in this paper. 

\subsection{The density matrix and its dynamics \label{Sec:QKE}}
At equilibrium, the density matrix for the Hamiltonian $\mH_0$ can be written in the energy band basis as follows, 
\begin{align}
 \rko=
 	\begin{pmatrix}
 	 n_1 & 0 & 0 & 0\\
 	 0 & n_2 & 0 & 0\\
 	 0 & 0 & n_3 & 0\\
 	 0 & 0 & 0 & n_4
 	\end{pmatrix}. 
\end{align}
At zero temperature, each of the distribution functions is a step function $n_i = \Theta(\varf-\eps{i})$, where $\varf$ is the Fermi energy. 

With Eq.~(\ref{Eq:QKE}) as our starting point,  $\rkl$ is again composed of two parts, $\rkl=\rklo+\rkle$, where $\rklo$ is the density matrix in the absence of interaction while $\rkle$ is the correction due to electron-electron interaction. 
Because of the $4\times 4$ matrix structure of Eq.~\eqref{Eq:QKE}, we introduce a complete set of 16 $\Gamma$ matrices [see Appendix~\ref{Appendix:GammaMatrices}] and expand the density matrix in the basis of these $\Gamma$ matrices. Specifically, the density matrix $\rkl$ is written as
\begin{align}
	\rkl=\sum_{i=1}^{16} \fkl_i \Gamma_i=\rklo+\rkle, \label{Eq:DensityMatrixExpansionFormal}
\end{align}
and the two terms are 
\begin{align}
	\rklo=\sum_{i=1}^{16} \fklo_i\Gamma_i,\quad \rkle=\sum_{i=1}^{16} \fkle_i\Gamma_i, \label{Eq:DensityMatrixExpansion}
\end{align}
where we have $\fkl_i=\fklo_i+\fkle_i$ for each $i$. In this way, the above matrix equation will be reduced to a set of coupled equations for these expansion coefficients. 

In addition, such a decomposition of the density matrix enables us to rewrite the current in Eq.~\eqref{Eq:totalcurrents} in a convenient form: the explicit expressions for $\bm{j}_{\bk}$ now read
\begin{align}
 j_k &\equiv \partii{\mH_0}{k} = \dfrac{\gamma_1}{\delk}\Gamma_1-\dfrac{2k}{\delk} \Gamma_{5}, \notag \\
 j_\phi &\equiv \dfrac{1}{k}\partii{\mH_0}{\phi} = \dfrac{-i\gamma_1}{\delk} \Gamma_7 +\dfrac{2k}{\delk} \Gamma_{14},
\end{align}
and the two currents $J_1$ and $J_2$ become
\begin{align}
	  J_1 &= 16e\int d\bk \cos\phi \left(\dfrac{\gamma_1}{\delk}\fkl_1-\dfrac{2k}{\delk}\fkl_{5}\right), \notag\\
	  J_2 &= 16e\int d\bk\sin\phi \left(\dfrac{i\gamma_1}{\delk}\fkl_{7}+\dfrac{2k}{\delk}\fkl_{14}\right).\label{Eq:TotalCurrents}
\end{align}
Note that only four expansion coefficients $\fkl_1$, $\fkl_5$, $\fkl_7$, and $\fkl_{14}$ contribute to the current. 

\subsection{Nonequilibrium density matrix $\rkl$ in the noninteracting limit \label{Sec:NonInteractingDensityMatrix}}
We first solve the nonequilibrium density matrix $\rkl$ in the absence of electron-electron interaction. Such a solution is obtained by using the noninteracting $\mathcal{H}_0$ [Eq.~\eqref{Eq:BilayerH0}] in the quantum kinetic equation [Eq.~\eqref{Eq:QKE}]. The resulting density matrix is just $\rklo$, according to our convention in Eq.~\eqref{Eq:DensityMatrixExpansion}. The corresponding 16 coefficients $\fklo_i$ are given below. 
First of all, four coefficients are proportional to the derivatives of the distribution functions: 
\begin{align}
	\begin{Bmatrix}
		\fklo_{5}\\ \fklo_{8}\\ \fklo_{9}\\ \fklo_{16}
	\end{Bmatrix}
	&=\dfrac{i\betaa(k)}{4\omega}
	\begin{Bmatrix}
		(n_1'-n_2'+n_3'-n_4')\\
		-i(n_1'-n_2'-n_3'+n_4')\\
		-(n_1'+n_2'-n_3'-n_4')\\
		-(n_1'+n_2'+n_3'+n_4')
	\end{Bmatrix}, \label{Eq:DeltaDistributionFunctions}
\end{align}
where the prime denotes partial derivatives, i.e., $n_i'(k)\equiv\partial n_i(k)/\partial k$. Secondly, we have the following coefficients, 
\allowdisplaybreaks[2]
\begin{align*}
	\begin{Bmatrix}
	\fklo_{6}\\ \fklo_{7}\\ \fklo_{10}\\ \fklo_{11}
	\end{Bmatrix}
	&=\dfrac{\gamma_1\betaa(k)}{2\delk^2(\delk^2-\omega^2)}
	\begin{Bmatrix}
		-\omega(n_1-n_2-n_3+n_4)\\
		-i\delk(n_1-n_2-n_3+n_4)\\
		\delk(n_1+n_2-n_3-n_4)\\
		i\omega(n_1+n_2-n_3-n_4)
	\end{Bmatrix},\notag\\
	\begin{Bmatrix}
	\fklo_{1}\\ \fklo_{4}\\ \fklo_{12}\\ \fklo_{15}
	\end{Bmatrix}
	&=\dfrac{\betab(k)}{2\delk(\gamma_1^2-\omega^2)}
	\begin{Bmatrix}
		\gamma_1 (n_1-n_2-n_3+n_4)\\
		\omega (n_1-n_2+n_3-n_4)\\
		\omega(n_1-n_2-n_3+n_4)\\
		\gamma_1(n_1-n_2+n_3-n_4)
	\end{Bmatrix},
\end{align*}
and finally,
\begin{align*}
	\fklo_{3}=&\dfrac{ik\betab(k)}{\delk}\left[\dfrac{n_1-n_4}{\omega^2-(\delk+\gamma_1)^2}+
			\dfrac{n_2-n_3}{\omega^2-(\delk-\gamma_1)^2}\right],\\
	\fklo_{13}=&\dfrac{k\betab(k)}{\delk}\left[\dfrac{n_1-n_4}{\omega^2-(\delk+\gamma_1)^2}-
				\dfrac{n_2-n_3}{\omega^2-(\delk-\gamma_1)^2}\right],\\
	\fklo_{2}=&\\
	\dfrac{\omega\betab(k)}{4k\delk}&\left[\dfrac{(n_1-n_4)(\delk-\gamma_1)}{\omega^2-(\delk+\gamma_1)^2}+\dfrac{(n_2-n_3)(\delk+\gamma_1)}{\omega^2-(\delk-\gamma_1)^2}\right],\\
	\fklo_{14}=&\\
	\dfrac{i\omega\betab(k)}{4k\delk}&\left[\dfrac{(n_1-n_4)(\delk-\gamma_1)}{\omega^2-(\delk+\gamma_1)^2}-\dfrac{(n_2-n_3)(\delk+\gamma_1)}{\omega^2-(\delk-\gamma_1)^2}\right].
\end{align*}
In the above expressions $\betaa(k)$ and $\betab(k)$ are given by 
\begin{align}
	\betaa(k)=\epa, \quad \betab (k)=\epb. \label{Eq:DefinitionOfBeta}
\end{align}
They represent two different couplings to the electric field. The total current in the noninteracting limit is then obtained by inserting the above results into the general equation Eq.~\eqref{Eq:TotalCurrents}. 

\subsection{Interaction corrections to the nonequilibrium density matrix \texorpdfstring{$\rkl$}{rho} \label{SubSection:Equations}} 
In the presence of electron-electron interaction, the nonequilibrium density matrix $\rkl$ will be further modified. The effect of interaction is incorporated by a quasiparticle exchange self-energy term in the Hamiltonian, which is given by
\begin{align}
	\Sigma(\bk) = -\sum_{\bk'}V_{\bk\bk'}\rho(\bk'),
\end{align}
where $V_{\bk\bk'}$ is the Coulomb potential. The property that the self-energy matrix at one wave vector is simply an interaction-weighted average of the density matrix at different wave vectors can be attributed to the model's pseudospin-independent interaction $V_{\bk\bk'}$. 

The quantum kinetic equation in Eq.~\eqref{Eq:QKE} now reads
\begin{align}
 -i\omega \rkl+e\bE\cdot\partii{\rko}{\bk}+i&\left[\mH_0, \rkl\right]+i\left[ \se(\bk), \rklo\right] \notag\\
 &=i\sum_{\bk'}V_{\bk\bk'}\left[\rklo(\bk'), \rko \right],\label{Eq:firstorderequation}
\end{align}
where $\se(\bk)$ is the equilibrium self-energy matrix,
\begin{align}
 \se(\bk) = -\sum_{\bk'} V_{\bk\bk'} \rko(\bk'). \label{Eq:Self-Energy}
\end{align} 
In the band basis, the diagonal entries of this matrix $\se_{\lambda}\equiv\bra{u_{\lambda}}\se\ket{u_{\lambda}}$ (where $\lambda=1,2,3,4)$ represent the electron self-energy in each band. In addition, in the presence of an applied electric field and electron-electron interaction, the quasiparticles are no longer in the same band eigenstates as the noninteracting electrons. As such, the equilibrium self-energy acquires off-diagonal entries in the band basis. We find that four off-diagonal self-energies are nonzero, $\se_{13}\equiv \bra{u_{1}(\bk)}\se\ket{u_{3}(\bk)}=(\se_{31})^{\ast}$, and $\se_{24}\equiv \bra{u_{2}(\bk)}\se\ket{u_{4}(\bk)}=(\se_{42})^{\ast}$, and the other entries vanish because of azimuthal symmetry. This is to be contrasted with the two-band model, where the self-energy matrix only contains diagonal entries~\cite{James:2009PRB_TwobandModel}. The explicit expressions for these matrix elements are given in Appendix~\ref{Appendix:SelfEnergy}. 

Directly substituting the expansion in Eq.~\eqref{Eq:DensityMatrixExpansionFormal} and rewriting the above matrix Eq.~\eqref{Eq:firstorderequation} yields a set of coupled 16 equations that is quite cumbersome and lacks transparency to their underlying physical meaning. We have found a better way to organize the coupled 16 equations with the following. 
We define a new set of variables from $\fkl_i$ as follows, $A_{\pm}=\fkl_2\pm i\fkl_{14}$, $B_{\pm}=i\fkl_3\pm \fkl_{13}$, $C_{\pm}=\fkl_{1}\pm \fkl_{15}$, $D_{\pm}=\fkl_{4}\pm \fkl_{12}$, $E_{\pm}=\fkl_{6}\pm i\fkl_{11}$, $F_{\pm}=\fkl_{10}\pm i\fkl_{7}$, $G_{\pm}=\fkl_{8}\pm i\fkl_{9}$, and $H_{\pm} =i\fkl_{5}\pm i\fkl_{16}$. With these new variables, the equations greatly simplify, and can be expressed as
\allowdisplaybreaks[2]
\begin{align}
\omega A_{+}+\delta_{23}B_{+}&-(n_2-n_3)e\bE\cdot\mathcal{A}_{23}\label{Eq:FirstTwoEq_1}\\&-\se_{13}C_{+}-\se_{24}C_{-}=\dse_{23,-},\notag\\
\delta_{23}A_{+}+\omega B_{+}&-\se_{13}D_{+}+\se_{24}D_{-}=\dse_{23,+},\label{Eq:FirstTwoEq_2}\\	 
\omega C_{+}+\delta_{21}D_{+}&-\se_{13}A_{+}-\se_{24}A_{-}=\dse_{21,+},\label{Eq:FirstTwoEq_3}\\
\delta_{21}C_{+}+\omega D_{+}&-(n_2-n_1)e\bE\cdot\mathcal{A}_{21} \label{Eq:FirstTwoEq_4} \\&-\se_{13}B_{+}+\se_{24}B_{-}=\dse_{21,-},\notag\\
\omega E_{+}+\delta_{13}F_{+}&-i(n_1-n_3)e\bE\cdot\mathcal{A}_{13}=\dse_{13,-},\label{Eq:FirstTwoEq_5}\\
\delta_{13}E_{+}+\omega F_{+}&-2\se_{13}G_{+}=\dse_{13,+},\label{Eq:FirstTwoEq_6}\\
\omega G_{+}&-2\se_{13}F_{+}=\epa (n_1'-n_3')/2,\label{Eq:FirstTwoEq_7}\\
\omega H_{+}&=\epa(n_2'+n_4')/2;\label{Eq:FirstTwoEq_8}\\
\notag\\
\omega A_{-}+\delta_{14}B_{-}&-(n_1-n_4)e\bE\cdot\mathcal{A}_{14}\label{Eq:FirstTwoEq_9}\\&-\se_{13}C_{-}-\se_{24}C_{+}=\dse_{14,-},\notag\\
\delta_{14}A_{-}+\omega B_{-}&-\se_{13}D_{-}+\se_{24}D_{+}=\dse_{14,+},\label{Eq:FirstTwoEq_10}\\	 
\omega C_{-}+\delta_{34}D_{-}&-\se_{13}A_{-}-\se_{24}A_{+}=\dse_{34,+},\label{Eq:FirstTwoEq_11}\\
\delta_{34}C_{-}+\omega D_{-}&
-(n_3-n_4)e\bE\cdot\mathcal{A}_{34}\label{Eq:FirstTwoEq_12}\\&-\se_{13}B_{-}+\se_{24}B_{+}=\dse_{34,-},\notag\\
\omega E_{-}+\delta_{42}F_{-}&-i(n_4-n_2)e\bE\cdot\mathcal{A}_{42}=\dse_{42,-},\label{Eq:FirstTwoEq_13}\\
\delta_{42}E_{-}+\omega F_{-}&-2\se_{24}G_{-}=\dse_{42,+},\label{Eq:FirstTwoEq_14}\\
\omega G_{-}&-2\se_{24}F_{-}=\epa(n_4'-n_2')/2,\label{Eq:FirstTwoEq_15}\\
\omega H_{-}&=-\epa(n_1'+n_3')/2.\label{Eq:LastEq}
\end{align}
where $\delta_{ij}=\eps{i}+\se_{i}-\eps{j}-\se_{j}$ is the energy needed to create a vertical interband excitation between band $i$ and $j$. The right-hand-side of Eqs.~\eqref{Eq:FirstTwoEq_1}-\eqref{Eq:LastEq} represent the nonequilibrium self-energy changes, whose detailed expressions are presented in Appendix~\ref{Appendix:EquationForG}. $\mathcal{A}_{ij}$ is the non-Abelian Berry connection defined in Eq.~\eqref{Eq:BerryConnection}.
This set of pseudospin Bloch equations generalize Eqs.~\eqref{Eq:Two-band-Bk}-\eqref{Eq:two-bandEquations} we obtained for the two-band Hamiltonian in Section~\ref{Sec:SLGDIC} to the four-band case. 
Eq.~\eqref{Eq:two-bandEquations} for the case of bilayer graphene ($m=2$) can be reproduced by Eqs.~\eqref{Eq:FirstTwoEq_1}-\eqref{Eq:FirstTwoEq_2} in the limit of large interlayer hopping energy ($\gamma_1\rightarrow\infty$), with $A_{+}(B_{+})$ in Eqs.~\eqref{Eq:FirstTwoEq_1}-\eqref{Eq:FirstTwoEq_2} given by 
$A_{+}\rightarrow P$ and $B_{+}\rightarrow Q$. 

Let us comment briefly on the physical meaning of the non-Abelian Berry connection appearing in our equations. 
It was shown in the context of semiclassical wavepacket dynamics~\cite{Yao:2005ke,DiXiao:2009RMP_BerryPhase} that such a coupling between the electric field $\bE$ and the non-Abelian Berry connection $\mathcal{A}_{ij}$ governs the redistribution of the electron occupations among different bands. These terms in our equations play a similar role. 
{To see this, note that }
such a coupling can be written explicitly as
\begin{align}
	e\bE\cdot \mathcal{A}_{ij}(\bk)&=i\betaa(k)\bra{u_i(\bk)} \frac{\partial}{\partial k} \ket{u_j(\bk)}\notag\\
	&\quad+i\betab(k)\bra{u_i(\bk)} \dfrac{1}{k}\frac{\partial}{\partial \phi} \ket{u_j(\bk)},
\end{align}
where $\betaa(k)$ and $\betab(k)$ are given by Eq.~\eqref{Eq:DefinitionOfBeta}. Interestingly, the six coupling terms in our equations fall naturally into two categories: $e\bE\cdot \mathcal{A}_{13}(\bk)$ and $e\bE\cdot \mathcal{A}_{42}(\bk)$ are proportional to $\betaa(k)$, while the other four couplings are proportional to $\betab(k)$. This correspondence is strikingly similar to the off-diagonal elements of the equilibrium self-energy matrix, where only $\se_{13}=(\se_{31})^{\ast}$ and $\se_{24}=(\se_{42})^{\ast}$ are nonzero [see Appendix~\ref{Appendix:SelfEnergy}]. 


We now explain the physical meaning of this set of coupled equations. The functions $G_{\pm}, H_{\pm}$ in Eqs.~(\ref{Eq:FirstTwoEq_7})-(\ref{Eq:FirstTwoEq_8}), (\ref{Eq:FirstTwoEq_15})-(\ref{Eq:LastEq}) describe Drude intraband dynamics for the four bands, whereas $A_{\pm}, B_{\pm}, C_{\pm}, D_{\pm}, E_{\pm}, F_{\pm}$ in other equations describe interband dynamics. Coupling between intraband and interband responses in these equations can be seen clearly as follows. 
We first note that the source terms $\bm{E}\cdot\bm{k}$ and $(\bm{E}\times\bm{k})_z$ in the kinetic equations respectively drive the  intraband and interband responses; the appearance of the Berry connection $\mathcal{A}_{13} \propto \bm{E}\cdot\bm{k}$ [$\mathcal{A}_{42}$] in Eqs.~\eqref{Eq:FirstTwoEq_5}-\eqref{Eq:FirstTwoEq_6} [\eqref{Eq:FirstTwoEq_13}-\eqref{Eq:FirstTwoEq_14}] therefore corresponds to a direct coupling of the interband transitions between bands $1$ and $3$ [$4$ and $2$] with the Drude intraband response. 
Due to exchange interaction, an indirect mechanism of Drude-interband coupling also occurs through the equilibrium $\Sigma^{(0)}_{13}$ [$\Sigma^{(0)}_{24}$] and nonequilibrium $\delta\Sigma^{(0)}_{13,+}$ [$\delta\Sigma^{(0)}_{42,+}$] self-energies. 
It is this interaction-induced Drude-interband coupling that gives rise to the renormalization of the optical Drude weight. 
The interband responses $A_{\pm}, B_{\pm}, C_{\pm}, D_{\pm}$ in Eqs.~(17)-(20), (25)-(28) couple to intraband responses only through the nonequilibrium self-energies $\delta\Sigma_{23,+}, \delta\Sigma_{21,+}, \delta\Sigma_{14,+}, \delta\Sigma_{34,+}$ through exchange effects.

These coupled equations can be solved numerically to yield $\rkle$, the interaction correction to the nonequilibrium density matrix $\rkl$. 
We then invert the equations to find the original coefficients $f_i$ and insert $\fkle_{1}$, $\fkle_{5}$, $\fkle_{7}$, and $\fkle_{14}$ into Eq.~\eqref{Eq:TotalCurrents} to obtain the interaction corrections to the optical conductivity and Drude weight. 

\section{Optical Drude weight in bilayer graphene \label{Sec:Drudeweight}}
In this section, we adopt the above formalism to obtain the optical Drude weight for bilayer graphene. We will first compute the optical conductivity in the noninteracting limit and show that it agrees with existing results. We will then turn on electron-electron interaction and study how it modifies the Drude weight. From now on, we will assume for concreteness that the Fermi energy $\varf>0$ is above the charge neutrality point. 

\subsection{Noninteracting results for the Drude weight }
In the noninteracting limit, the optical conductivity of bilayer graphene is obtained by inserting the noninteracting density matrix $\rklo$ found in Section~\ref{Sec:NonInteractingDensityMatrix} into Eq.~\eqref{Eq:TotalCurrents}. As a result, the real and imaginary parts of the conductivity are given explicitly by
\begin{widetext}
\begin{align}
 \dfrac{\text{Re}[\sigma(\Omega)]}{\sigma_0} &=
 \dfrac{\Theta(\Omega-1)}{4 \Omega^2}[\Theta(\Omega-2\mu-1)+\Theta(\Omega-2\mu+1)]
 +\Theta(\Omega-2\mu)\left[\dfrac{\Omega+2}{4(\Omega+1)} +\Theta(\Omega-2)\dfrac{\Omega-2}{4(\Omega-1)} \right]\notag\\
 &\quad +\drudeo_1 \delta(\Omega)+\drudeo_2\delta(\Omega-1),\notag\\
 \dfrac{\text{Im}[\sigma(\Omega)]}{\sigma_0/\pi} &=
   \dfrac{1}{4\Omega^2}\left[\ln\left|\dfrac{1-\Omega}{1+\Omega}\right| \Theta(1-\mu)
  +\left(\ln\left|\dfrac{1+2\mu-\Omega}{1+2\mu+\Omega}\right| +\dfrac{4\Omega(\mu+1)}{2\mu+1}\right)
  -\Theta(\mu-1)\left(\ln\left|\dfrac{1-2\mu-\Omega}{1-2\mu+\Omega}\right|+\dfrac{4\Omega(\mu-1)}{2\mu-1}\right)\right]\notag\\
  &\quad+\dfrac{\drudeo_1}{\Omega}+\dfrac{2\Omega\drudeo_2}{\Omega^2-1}-\dfrac{1}{4}\big[r(\Omega,1)-r(\Omega,-\mu)+\Theta(\mu-1)(r(\Omega,\mu)-r(\Omega,1))\big], 
  \label{Eq:Conductivity}
\end{align}
\end{widetext}
where $\sigma_0=e^2/\hbar$. In the above results, $\mu\equiv\varf/\gamma_1$ and $\Omega\equiv \omega/\gamma_1$ are the Fermi energy and optical frequency normalized by interlayer hopping energy $\gamma_1$, respectively. In addition,  the function $r(\Omega,\mu)$ is given by
\begin{align}
 r(\Omega,\mu)=
  \dfrac{\Omega+2}{\Omega+1}\ln|\Omega+2 \mu| -\dfrac{\Omega-2}{\Omega-1}\ln|\Omega-2 \mu|\notag\\
 -\dfrac{2 \Omega}{\Omega^2-1}\ln|2 \mu-1|. \label{Eq:r-function}
\end{align}
Finally, the coefficients in front of the two delta functions are the optical weight for the $\omega=0$ and $\omega=\gamma_1$ peak, respectively, 
\begin{align}
 \drudeo_1(\mu) &= \dfrac{2\mu(\mu+1)}{2\mu+1}+\dfrac{2\mu(\mu-1)}{2\mu-1}\Theta(\mu-1),\notag\\
 \drudeo_2(\mu) &= \dfrac{1}{4}[\ln(2\mu+1)-\Theta(\mu-1)\ln(2\mu-1)].\label{Eq:DrudeZero}
\end{align}
This result agrees with previous studies~\cite{Nicol:2008PRB_BLGCond,Nilsson:2008PRB_ElePropertyBLG,Zhang:2008PRB_BLGconductivity,Abergel:2007PRB_OpticalInfrared,McCann:2007SSCom_ElectronsBLG,Benfatto:2008PRB_OpticalSumRule}, and has been discussed extensively in the literature. Here we just want to emphasize that only $\drudeo_1$ arises from intraband contributions, and is the Drude weight we are looking for. The $\drudeo_2$ peak $\omega=\gamma_1$ arises from optical transitions between the two conduction bands, and its delta function dependence is due to the constant energy difference $\gamma_1$ between the two bands in our model. 
When a band gap is opened~\cite{Nicol:2008PRB_BLGCond} or remote hopping parameters are taken into account~\cite{Wright:2009Nanotechnology_EffectOfNNNhopping}, 
the two bands will no longer be energetically equidistant and 
 the sharp peak at $\omega=\gamma_1$ will then be broadened.  

\subsection{Interaction corrections to the Drude weight \label{Sec:CondInteraction}}
Now we are going to study how electron-electron interaction modifies the Drude weight in bilayer graphene. In general,  the Eqs.~\eqref{Eq:FirstTwoEq_1}-\eqref{Eq:LastEq} can be solved numerically to obtain the nonequilibrium density matrix $\rkl$ to all orders of interaction potential. However, this is quite complicated, and we will introduce some simplifications. 

First, we will only solve these coupled equations perturbatively and obtain lowest order interaction corrections to the Drude weight. Therefore, we will keep terms up to first order in the interaction potential in these equations, which allows us to obtain closed-form solutions for the coefficients $\fkle_1$, $\fkle_5$, $\fkle_7$, and $\fkle_{14}$ as follows, 
\begin{widetext}
\begin{align}
 	\fkle_{1} &= \dfrac{i\gamma_1(\dse_{21,-}-\dse_{34,-})+i\omega(\dse_{21,+}+\dse_{34,+})}{2(\omega^2-\gamma_1^2)}, \; \fkle_{7} = \dfrac{-i\Delta_{\bk} (\dse_{13,-}+\dse_{42,-})-i\omega (\dse_{42,+}-\dse_{13,+})}{2(\Delta_{\bk}^2-\gamma_1^2)}, \; \fkle_{5}=0, \notag\\
	\fkle_{14}&=[2(\omega^2-(\delk+\gamma_1)^2)(\omega^2-(\gamma_1-\delk)^2)]^{-1} \Big[2i\omega\gamma_1\delk(\dse_{21,-}-\dse_{34,-})
	+i\gamma_1(\gamma_1^2-\delk^2-\omega^2)(\dse_{23,+}+\dse_{14,+})\notag\\
	&-i\delk(\gamma_1^2+\omega^2-\delk^2)(\dse_{23,+}-\dse_{14,+})
	+i\omega(\delk^2+\gamma_1^2-\omega^2)(\dse_{14,-}-\dse_{23,-})\Big]
	.\label{Eq:F14}
\end{align}
\end{widetext}
Here, to first order in the interaction potential, the nonequilibrium self-energy changes in the above expression are given by Appendix~\ref{Appendix:EquationForG} with $A_{\pm}$ to $H_{\pm}$ taking their noninteracting values given in Sections~\ref{Sec:NonInteractingDensityMatrix}-\ref{SubSection:Equations}. 

In addition, because we are only concerned with the Drude weight, it is sufficient to extract the $\omega^{-1}$ dependence in these coefficients. We then note that $\fkle_5$ vanishes, and hence does not contribute to the conductivity. In addition, neither $\fkle_1$ nor $\fkle_7$ contains an overall $\omega^{-1}$ dependence. Therefore, only the $\fkle_{14}$ term matters. We can then extract the coefficient in front of $\omega^{-1}$ as 
\allowdisplaybreaks[2]
\begin{align}
	\fkle_{14}\sim &\Big\{(n_1-n_2+n_3-n_4)\delk\left[2\gamma_1(S_1+S_3)+8kS_2\right]\notag\\
	 -(n_1+n_2&-n_3-n_4)\left[(\delk^2+\gamma_1^2)(S_1+S_3)+8k\gamma_1S_2\right]\Big\}\notag\\
	 &\times(16\delk k^2)^{-1}, \label{Eq:G14}
\end{align}
where the functions $S_1(k)$, $S_2(k)$ and $S_3(k)$ are
\begin{align}
 S_1(k) &=\sum_{\bk'}V_{\bk\bk'} \fklo_{5}(\bk')\sin2\fkk,\notag\\
 S_2(k) &= \sum_{\bk'}V_{\bk\bk'} \fklo_{9}(\bk')\sin\fkk\dfrac{k'}{\delkp},\notag\\
 S_3(k) &=\sum_{\bk'}V_{\bk\bk'} \fklo_{9}(\bk')\sin2\fkk \dfrac{\gamma_1}{\delkp}, \label{Eq:S1S2S3}
\end{align}
and $\fkk$ is the angle between momenta $\bk$ and $\bk'$. 
For future convenience, we also define 
\begin{align}
	s_{j}=16\pi S_{j}/ie^2\betab(k), \;(j=1,2,3).\label{Eq:s1s2s3}
\end{align}
Note that the $\omega^{-1}$ dependence of $\fkle_{14}$ comes from $\fklo_{5}$ and $\fklo_{9}$ only [see Eq.~\eqref{Eq:DeltaDistributionFunctions}]. 
The explicit expressions for $s_1(k)$, $s_2(k)$ and $s_3(k)$ will depend on the form of Coulomb potential we adopt in the calculation. Other than this choice, Eqs.~\eqref{Eq:G14}-\eqref{Eq:s1s2s3} represent the most general form for the $\omega^{-1}$ dependence in $\fkle_{14}$ and hence $\rkle$, which can then be inserted into Eq.~\eqref{Eq:TotalCurrents} to obtain its contribution to the conductivity as follows
\begin{align}
 \sigma^{(e)} &=  \dfrac{ie^4\gamma_1}{32\pi^2\omega}\int_0^{\tdk_c}d\tdk\dfrac{\mathcal{G}(\tdk)}{4\tdk^2+1}.
 \label{Eq:Sigmae}
\end{align}
Here the integration cutoff is set by the Brillouin zone boundary $ k_c=1/a=\SI{7.0e9}{\per\m}$, where $a=\SI{1.43}{\angstrom}$ is the carbon-carbon distance. This corresponds to an energy scale of $\Lambda=\hbar vk_c\approx 11.50\gamma_1$. 
Also, we have changed the integration variable to a dimensionless one $\tdk\equiv \hbar vk/\gamma_1$, and similarly $\tdk_c\equiv\hbar vk_c/\gamma_1$. The function $\mathcal{G}(\tdk)$ in the integrand is given by
\begin{align}
\mathcal{G}(k)&=
	\omega\big[8k   s_2(k)+(4k^2+2)( s_1(k)+ s_3(k))\big]\notag\\
	&\quad\times(n_1+n_2-n_3-n_4)\notag\\
	 &-\omega\sqrt{4k^2+1}\big[2( s_1(k)+ s_3(k) )+8k s_2(k)\big]\notag\\
	 &\quad\times (n_1-n_2+n_3-n_4).
\end{align}
Because the functions $s_j (j=1,2,3)$ are all proportional to $\omega^{-1}$ by virtue of Eqs.~\eqref{Eq:DeltaDistributionFunctions},~\eqref{Eq:S1S2S3},~and \eqref{Eq:s1s2s3}, $\mathcal{G}(\tdk)$ is independent of $\omega$. If we then replace $\omega^{-1}$ by $-i\delta(\omega)$ and restore $\hbar$ and $v$ in  Eq.~\eqref{Eq:Sigmae}, the leading order interaction correction to the Drude weight now reads
\begin{align}
 \drudee_1= \dfrac{e^2}{\hbar}\dfrac{\alpha^\ast\gamma_1}{32\pi}\int_0^{\tdk_c}d\tdk\dfrac{\mathcal{G}(\tdk)}{4\tdk^2+1}. \label{Eq:Drude-2}
\end{align}

It was shown in Ref.~\onlinecite{James:2009PRB_TwobandModel} that broken Galilean invariance gives rise to a peculiar mechanism that couples the Drude response to the interband response in bilayer graphene. This is the very reason why electron-electron interaction can modify the Drude weight in bilayer graphene. The importance of such a coupling can be quantified by the interaction-induced Drude weight renormalization $\dic$,
\begin{align}
 \dic-1= \dfrac{\drudee_1}{\drudeo_1} = 
 \dfrac{\displaystyle\dfrac{\alpha^{\ast}}{32\pi\mu}\int_0^{\tdk_c}d\tdk\dfrac{\mathcal{G}(\tdk)}{4\tdk^2+1}}
 {\dfrac{2(\mu+1)}{2\mu+1}+\dfrac{2(\mu-1)}{2\mu-1}\Theta(\mu-1)}. \label{Eq:DIC-General}
\end{align}

The rest of the section will be devoted to calculations of $\dic$. Before presenting the results, however, we note that in this work we will only include static screening effects, and use the following Coulomb potential
\begin{align}
 V_{\bk\bk'}= \dfrac{2\pi e^2}{|\bk-\bk'|+\ktf},  \label{Eq:CoulombPotential}
\end{align}
where $\ktf$ is the Thomas-Fermi screening length [see Appendix~\ref{Appendix:Thomas-Fermi} for derivations] given by
\begin{align}
 \ktf = \dfrac{2\alpha^{\ast}}{v} \left[2\varf+\gamma_1+\Theta(\varf-\gamma_1)(2\varf-\gamma_1)\right].\label{Eq:TFScreening}
\end{align}
Also note that there is a discontinuity in $\ktf$ at $\varf=\gamma_1$, i.e., when the Fermi energy moves into the higher conduction band. In the limit of small Fermi energy $\varf\ll \gamma_1$, this result correctly reduces to $2\alpha^{\ast}\gamma_1/v$, the result deduced from a two-band model of bilayer graphene~\cite{DasSarma:2011RMP_GrapheneTransport}. 

In what follows, we first consider two limits where analytical expressions for $\dic$ can be obtained. When the interaction is weakly screened, we can neglect the $\ktf$ term in the Coulomb potential [see Eq.~\eqref{Eq:CoulombPotential}]. This corresponds to the limit of long-range interaction. In contrast, when the interaction is heavily screened, the Thomas-Fermi screening length $\ktf$ will be much larger than typical values of $|\bk-\bk'|$. 
Thus, we can keep only the $\ktf$ term in the denominator of the Coulomb potential, which corresponds to the limit of short-range interaction. Finally, we will compare these two results to exact numerical evaluations of Eq.~\eqref{Eq:DIC-General} using the full Coulomb potential in Eq.~\eqref{Eq:CoulombPotential}. 

\subsubsection{Long-range interaction limit}
In this limit, we ignore screening effects entirely and set $\ktf$ in Eq.~\eqref{Eq:CoulombPotential} to zero. We then obtain analytical expressions for the functions $s_i$ in Eq.~\eqref{Eq:S1S2S3} as follows
 \begin{align}
 \allowdisplaybreaks
  s_1(k) &= \dfrac{\kbig}{\omega}[\Phi_3(k,\kbig)-\Phi_1(k,\kbig)]\notag\\
  &\quad-\dfrac{\ksmall}{\omega}\Theta(\varf-\gamma_1)[\Phi_3(k,\ksmall)-\Phi_1(k,\ksmall)], \notag\\
  s_2(k) &= \dfrac{\kbig^2}{\omega\Delta_{+}}[\Phi_2(k,\kbig)-\Phi_0(k,\kbig)]\notag\\
         &\quad+\dfrac{\ksmall^2\Theta(\varf-\gamma_1)}{\omega\Delta_{-}}[\Phi_2(k,\ksmall)-\Phi_0(k,\ksmall)],\notag\\
  s_3(k) &= \dfrac{\gamma_1\kbig}{\omega\Delta_{+}}[\Phi_3(k,\kbig)-\Phi_1(k,\kbig)] \label{Eq:S1S2S3ForLongRangeCoulomb}\\
         &\quad+\dfrac{\gamma_1\ksmall\Theta(\varf-\gamma_1)}{\omega\Delta_{-}}[\Phi_3(k,\ksmall)-\Phi_1(k,\ksmall)].\notag
 \end{align}
where $\Delta_{\pm}=\sqrt{4k_{\pm}^2+\gamma_1^2}$. In addition, $\kbig$ ($\ksmall$) is the Fermi wave-vector at which the Fermi energy intersects with the lower (upper) conduction band, defined as
\begin{align}
 \ksmall &= \Theta(\varf-\gamma_1)\sqrt{\varf(\varf-\gamma_1)}, \notag\\ \kbig&=\sqrt{\varf(\varf+\gamma_1)}.\label{Eq:FermiWaveVector}
\end{align}
The special functions $\Phi_i(k,k')$ arise from the integration of the long-range Coulomb potential over $\fkk$, and their expressions are given in Eq.~\eqref{Eq:AngularIntegrals}. 

Before the numerical evaluation of $\dic$, we want to show that our result correctly reduces to the one obtained by a two-band model of bilayer graphene~\cite{James:2009PRB_TwobandModel} in the $\varf\ll\gamma_1$ limit, which proceeds as follows. 
First, in the limit of $\varf\rightarrow 0$, the two Fermi wave-vectors satisfy $\tdksmall\equiv \hbar v\ksmall/\gamma_1\rightarrow 0$ and $\tdkbig\equiv\hbar v\kbig/\gamma_1\simeq\sqrt{\varf/\gamma_1}=\sqrt{\mu}$. The three special functions $s_i(k)$ in this limit thus satisfy $s_1 = \gamma_1s_3$ and $s_2=0$. In addition, $\delk$ can be approximated by $\gamma_1$. As a result, the integrand of Eq.~\eqref{Eq:DIC-General} reduces to 
\begin{align}
  \dfrac{\mathcal{G}(\tdk)}{4\tdk^2+1}\rightarrow -8\omega s_1(\tdk)\Theta(\tdk-\tdkbig).
\end{align}
We further note that the function $s_1(k)$ in the limit of $\varf\ll \gamma_1$ can be written as 
\begin{align*}
 \omega s_1(\tdk) \equiv \tdkbig[\Phi_3(\tdk,\tdkbig)-\Phi_{1}(\tdk,\tdkbig)]=-4\mathcal{R}\left(\dfrac{\tdk}{\tdkbig}\right), 
\end{align*}
where the function $\mathcal{R}(y)$ is given by
\begin{align}
 \mathcal{R}(y)= \dfrac{4(y+1)(y^4-y^2+1)}{15y^3}\elle\left(\dfrac{4y}{(y+1)^2}\right)\notag\\
 -\dfrac{4(y^2+1)(y-1)^2(y+1)}{15y^3}\ellk\left(\dfrac{4y}{(y+1)^2}\right),
\end{align}
and $\ellk(z)$ [$\elle(z)$] is the complete elliptic integrals of the first (second) kind [see Eq.~\eqref{Eq:EllipticIntegrals}]. Finally, in the limit of $\mu\ll 1$, the denominator in Eq.~\eqref{Eq:DIC-General} reduces to a constant 2. Putting everything together we can obtain $\dic$ in the limit $\mu\ll 1$ as follows, 
\begin{align}
 \bar{\mathcal{D}}-1 = \dfrac{1}{2}\dfrac{\alpha^{\ast}}{\pi\mu}\int_{\sqrt{\mu}}^{\tdk_c}d\tdk\; 32\mathcal{R}(\tdk)
 =\dfrac{\alpha^{\ast}\displaystyle\int_1^{\frac{\tdk_c}{\sqrt{\mu}}}\mathcal{R}(y) dy}{2\pi\sqrt{\mu}}.\label{Eq:Two-band}
\end{align}
This result agrees with the one obtained in Ref.~\onlinecite{James:2009PRB_TwobandModel}, which confirms the validity of our theory in the $\mu\ll 1$ limit. 

\begin{figure}[!]
\centering
\includegraphics[scale=0.5]{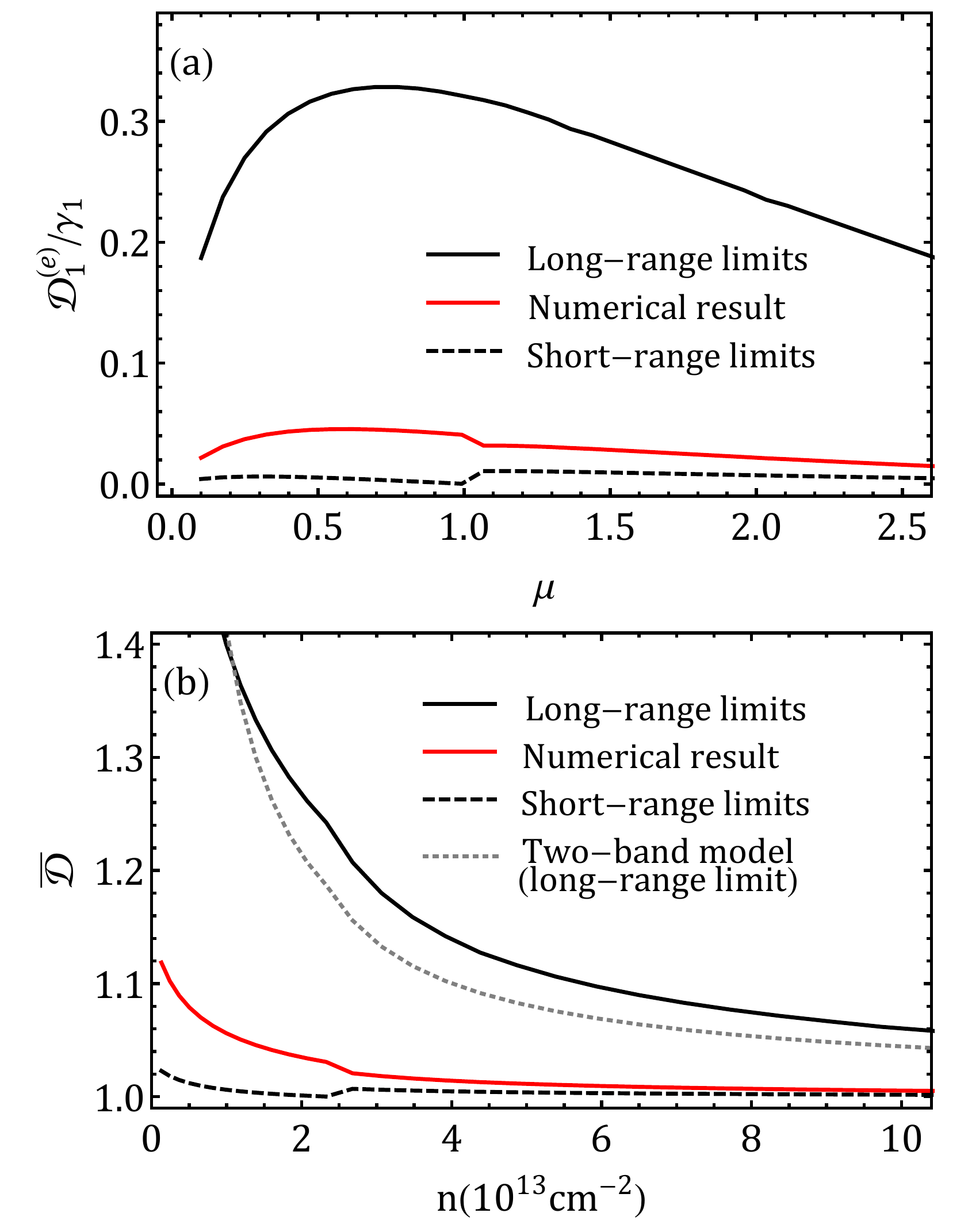}
\caption{(a) Interaction corrections to the Drude weight in bilayer graphene [see Eq.~\eqref{Eq:Drude-2}]. We compare the long-range limit (black solid line), short-range limit (dashed line), and the numerical result (red solid line). Here $\mu\equiv \varepsilon_F/\gamma_1$, and dielectric constant $\kappa=1$. The discontinuity in the short-range and numerical results is due to the discontinuity of $\ktf$ at $\mu=1$ (see the discussions in Appendix~\ref{Appendix:Thomas-Fermi}). (b) Interaction-induced Drude weight renormalization $\dic$ [see Eq.~\eqref{Eq:DIC-General}]. As a comparison, we also show the leading-order (in interaction strength) $\dic$ in the long-range limit obtained previously by the two-band model (gray dotted line) in Ref.~\onlinecite{James:2009PRB_TwobandModel}. Note that the leading-order short-range $\dic$ in the two-band model vanishes. \label{Fig:DIC}}
\end{figure}

We now evaluate $\dic$ in the long-range interaction limit, which is shown in the black solid line in Fig.~\ref{Fig:DIC}. Several comments are in order. 
First, the Drude weight correction $\dic-1\sim10\%-40\%$ [Fig.~\ref{Fig:DIC}(b)], which indicates that the interaction correction to the Drude weight is at least one order of magnitude smaller than the noninteracting Drude weight. 
This shows that our perturbative solutions to the quantum kinetic equation is well controlled, even the expansion parameter $\alpha^{\ast}\equiv e^2/\kappa v$ may not be small. 
In addition, the evaluation of $\dic$ in this long-range limit yields a convergent result, in contrast to the short-range limit, which has a logarithmic dependence on the cutoff $\Lambda$ [see Eq.~\eqref{Eq:Short-range} below]. 
In contrast, $\dic$ in single-layer graphene has a logarithmic dependence on the cutoff $\Lambda$ in the long-range limit and a linear dependence in the short-range limit~\cite{Polini:2011PRB_Drudeweight}. 
Finally, although the Drude weight renormalization $\dic$ in our theory can be reduced to the one obtained by the two-band model in the limit of $\mu\ll1$, the latter tends to underestimate the interaction corrections when the electron density is higher [Fig.~\ref{Fig:DIC}(b)]. This is expected because the analysis based on the two-band model cannot account for the contributions from the higher energy bands $\varepsilon_1$ and $\varepsilon_4$. 

\subsubsection{Short-range interaction limit}
We now consider the opposite limit where the electron-electron interaction is heavily screened {and hence effectively short-ranged}, so that we can ignore the momentum dependence in the Coulomb potential in Eq.~\eqref{Eq:CoulombPotential}. Therefore, $V_{\bk\bk'}$ is now a constant and no longer subject to the angular integration over $\fkk$. As a result, both $s_1(k)$ and $s_3(k)$ vanish, while $s_2(k)$ becomes
\begin{align}
 s_2(k)=-2\pi\left[\dfrac{\tdkbig}{\tilde{\Delta}_{+}}\dfrac{\kbig}{\ktf}+\Theta(\varf-\gamma_1)\dfrac{\tdksmall}{\tilde{\Delta}_{-}}\dfrac{\ksmall}{\ktf}\right],
\end{align}
where $\tdk_{F\pm}\equiv\hbar vk_{F\pm}/\gamma_1$ and $\tilde{\Delta}_{\pm}\equiv \sqrt{4\tdk_{F\pm}^2+1}$. The $\dic$ in this limit is given by
\begin{align}
\allowdisplaybreaks[4]
\dic-& 1 =\dfrac{\alpha^{\ast}\gamma_1\left[\dfrac{\tdkbig}{\tilde{\Delta}_{+}}\dfrac{\kbig}{\ktf}+\Theta(\mu-1)\dfrac{\tdksmall}{\tilde{\Delta}_{-}}\dfrac{\ksmall}{\ktf}\right]}{16\pi\left[\dfrac{2\mu(\mu+1)}{2\mu+1}+\dfrac{2\mu(\mu-1)}{2\mu-1}\Theta(\mu-1)\right]}\label{Eq:Short-range}\\
\times \Big[& \ln\dfrac{4\tdk_c^2+1}{4\tdkbig^2+1}-\Theta(\mu-1)\ln\dfrac{4\tdk_c^2+1}{4\tdksmall^2+1}\notag\\
 -2&\big(\sqrt{4\tdkbig^2+1}-1\big)+2\Theta(\mu-1)\big(\sqrt{4\tdksmall^2+1}-1\big)\Big].\notag
\end{align}
where $\Lambda$ is the energy cutoff introduced in Eq.~\eqref{Eq:Drude-2}. In the limit that the Fermi energy is much higher than the bottom of the higher conduction band ($\varf\gg\gamma_1$), this result can be simplified to
\begin{align}
	\dic-1 \simeq \dfrac{\gamma_1}{64\pi\varf}\left[\ln(\hbar vk_c/\varf)-1\right], \quad \varf\gg\gamma_1.
\end{align}
We find that this result approximates the full expression in Eq.~\eqref{Eq:Short-range} within 1\%  when $\varf\geq 2\gamma_1$. 

The short-range result in Eq.~\eqref{Eq:Short-range} is shown by the black dashed line in Fig.~\ref{Fig:DIC}. Several comments are in order. First, this result is much smaller than $\dic$ in the long-range limit. This is because the Thomas-Fermi screening length $\ktf$ is actually fairly large in bilayer graphene [see Eq.~\eqref{Eq:TFScreening}]. In fact, when the Fermi energy exceeds $\gamma_1$, $\ktf$ can be much larger than the momentum cutoff $\Lambda$ introduced in Eq.~\eqref{Eq:Short-range}. Secondly, we note that this is a new result that cannot be obtained by the two-band model of bilayer graphene, as Ref.~\onlinecite{James:2009PRB_TwobandModel} predicts a vanishing $\dic$ in this short-range interaction limit. This further suggests that interaction corrections to the optical conductivity may not vanish even in the short-range limit. Finally, the short-range $\dic$ in Eq.~\eqref{Eq:Short-range} is actually independent of the effective fine-structure constant $\alpha^{\ast}$, as the $\alpha^{\ast}$ in the numerator will cancel the $\alpha^{\ast}$ dependence in $\ktf$ in the denominator. This suggests that this short-range result is not affected by the interaction strength of the system. 

\subsubsection{Comparison with numerical results}
Having studied the interaction corrections to the Drude weight in the above two limits, we now compare them with numerical evaluations [red solid lines in Fig.~\ref{Fig:DIC}]. 
We note that the short-range limit [Eq.~\eqref{Eq:Short-range}] gives a very good approximation to the numerical result. 
To shed light on this result, note that the Thomas-Fermi screening wavevector $\ktf$ is extremely large {in bilayer graphene}. From Eq.~\eqref{Eq:TFScreening}, we can see that the $\ktf$ does not vanish even when the Fermi energy is at the charge neutrality point $\varf=0$. This reflects the constant density of states in bilayer graphene, even at the charge neutrality point. 
We therefore find the lower bound for $\ktf$ to be $2\alpha^\ast\gamma_1/\kappa v=\SI{2.65e9}{\per\cm}$. Such a momentum corresponds to a band energy of about $4\gamma_1$, which is four times the interlayer coupling energy. 
The large screening wavevector thus makes the electron-electron interaction effectively short-ranged, which explains why the numerical calculation of $\dic$ can be  approximated reasonably well by the short-range limit.

\section{Discussions \label{Sec:Discussions}}
In regular semiconductors with a parabolic band dispersion, Galilean invariance also prevents plasmon frequency $\omega_p$ from long-wavelength interaction renormalization~\cite{Kohn:1961_PR_CyclotronResonance}. One may wonder whether $\omega_p$ is also modified in graphene. We argue that because the Drude weight is closely related to the plasmon frequency, the latter should also be modified in graphene. To show this, we start 
from the well-known relation between the conductivity and the polarizability, 
\begin{align}
	\sigma(\omega) = \lim_{q\rightarrow 0} \dfrac{ie^2\omega}{q^2}\Pi(q,\omega),
\end{align}
which indicates that the real part of the polarizability in the limit of $vq\ll \omega$ and $\omega\ll \varf$ is given by 
\begin{align}
	\text{Re}\,\Pi(q,\omega) = \dfrac{\gamma_1 \tilde{\mathcal{D}}_1}{\pi}\left(\dfrac{q}{\omega}\right)^2.
\end{align}
The renormalized plasmon frequency $\omega_p$ is thus given by the {zero} of the dielectric function $\epsilon(q,\omega)=1-V(q)\text{Re}\Pi(q,\omega)=0$. For bilayer graphene we find that 
\begin{align}
	\omega_p^2 = 2e^2\gamma_1\tilde{\mathcal{D}}_1 q.
\end{align}
One can see immediately that the plasmon frequency $\omega_p^2$ is directly proportional to $\dic$, the Drude weight renormalization. 
Our prediction of interaction-modified plasmon frequency can be verified experimentally by using electron energy-loss spectroscopy of suspended bilayer graphene samples. 

We mention that the theory presented in this work 
is applicable for Fermi energies $\varf>\varepsilon_{c}\equiv \gamma_1(\gamma_3/\gamma_0)^2/2\sim \SI{10}{\meV}$~\cite{McCann:2006ev}. 
For {a smaller carrier density}, the effects of trigonal warping and electron-hole asymmetry become non-negligible~\cite{Malard:2007PRB_Raman,Zhang:2008PRB_BLGconductivity,Kuzmenko:2009PRB_InfraredSpect,Li:2009PRL_BandAsymmetry,Varlet2014:BLG} and can be easily incorporated into our theory through the Hamiltonian in Eq.~\eqref{Eq:BilayerH0}. 
We expect such effects will give rise to quantitative differences in the renormalized Drude weight, but do not alter our main qualitative conclusions. 

In addition, we wish to emphasize that some effects can only be captured by the full four-band model but not the two-band model, even at relatively low doping levels $\varf<\gamma_1$. The reason is two-fold. First of all, our full four-band calculation can capture the additional interband transitions involving the higher conduction and lower valence bands, as well as the transitions between the two conduction (valence) bands. These ingredients cannot be included in the two-band treatments~\cite{James:2009PRB_TwobandModel}. Secondly, even the low energy bands are not well captured in the two-band description of bilayer graphene, because the dispersion quickly deviates from being parabolic when $\varf\gtrsim \gamma_1/4$~\cite{McCann:2006ev}. Indeed, our calculations show that these effects give rise to important differences. For example, in the long-range limit the interaction corrections to the Drude weight are qualitatively different in both cases. In addition, in the short-range limit the interaction corrections to the Drude weight completely vanish in the two-band calculation, while we find finite corrections in the four-band treatments [see Fig.~\ref{Fig:DIC}(b)]. 

One of the interesting properties of bilayer and multilayer graphene systems is the opening of a band gap achieved by breaking the symmetry of the layer degrees of freedom with an applied out-of-plane voltage. We expect that the effects of Drude weight and plasmon frequency renormalization to be suppressed by a band gap. The physical reason is that  the renormalization effects arise from the coupling between the interband and the Drude intraband responses. Such a coupling is diminished by an increasing value of the band gap as the Dirac sea of valence-band electrons at $k > k_F$ moves farther apart from the conduction-band Fermi surface. Another way to look at this is by noting that the pseudospins become increasingly aligned with the out-of-plane $z$ direction with an increasing band gap. The pseudospin texture of all states therefore become more uniform, with the pseudospin of each quantum state becoming more similar. When an external electric field is applied under transport conditions, the degree to which the Galilean invariance symmetry is broken will be less severe, suppressing the interaction-induced renormalization effects.


\section{Conclusions\label{Sec:Conclusion}}

In this paper we have developed a theory for the optical conductivity of chiral multilayer graphene based on a quantum kinetic approach including the effects of electron-electron interaction. 
Our theory is first applied to the two-band model of chiral multilayer graphene and then generalized to the four-band model of bilayer graphene. We have obtained the equations of motion for the pseudospin components of the density matrix, which generalizes the semiconductor Bloch equation in conventional  parabolic band electron systems to chiral electron systems with pseudospins. 
From these equations we have calculated the interaction-induced corrections to the optical Drude weight, quantified by the Drude weight renormalization $\dic$. We find that $\dic$ increases with decreasing number of layers and hence pseudospin winding number, reaching the largest value in single-layer graphene. $\dic$ is also found to increase with decreasing electron density. Finally, we note that the renormalization effects of Drude weight and plasmon frequency are not limited to graphene systems. Our work has direct implications on the optical properties of other materials whose electronic states are also chiral or helical. In many topological states of matter, Galilean invariance of electronic states near the nodal points is  explicitly broken due to the helicity of the low-energy electrons. As a result, we expect interaction-induced renormalization effects of Drude weight and plasmon frequency also in chiral systems such as monolayer MoS$_2$~\cite{Xiao:2012dv,Li:2013tx,Tianyi_MagneticControl}, topological insulators~\cite{Hao:2010ku,Qi:2011hb,Vafek2014,Qiao_BLG}, and topological crystalline insulators~\cite{Hsieh:2012NatCommn_SnTe,Dziawa:2012NatMater_PbSnSe,Tanaka:2012NatPhys_SnTe,chiu2016type-II}.

\begin{acknowledgments}
We are greatly indebted to A.~H.~MacDonald and M.~Polini, who shared with us many insightful discussions. 
X. L. was supported by the U.S. DOE (Grant No. DE-FG03- 02ER45958, Division of Materials Science and Engineering) and the Welch Foundation (Grant No. F- 1255) in Austin, Texas, and is currently supported by JQI- NSF-PFC and LPS-MPO-CMTC in Maryland.
W.-K. is supported by a startup fund from the University of Alabama. 
\end{acknowledgments}

\appendix
\section{The unitary transformation that diagonalizes the noninteracting Hamiltonian~\label{Appendix:GammaMatrices}}
In our discussions of the two-band models, we introduced a set of generalized Pauli matrices in Eq.~\eqref{Eq:TwobandRk}. In fact they are obtained from a unitary transformation, 
\begin{align}
	\tilde{\sigma}_i = \mathcal{M}\sigma_i \mathcal{M}^{\dagger}, 
\end{align}
where $\mathcal{M}$ is the unitary transformation that diagonalizes the two-band Hamiltonian $\mathcal{H}_0 = \epsilon_{\bk}\hat{\bm{n}}\cdot\bm{\sigma}$, given by 
\begin{align}
	\mathcal{M} = \dfrac{1}{\sqrt{2}}
\begin{pmatrix}
	e^{-il\phi} & e^{-il\phi}\\ -1 & 1
\end{pmatrix}.
\end{align}
The explicit expressions for the generalized Pauli matrices are the following 
\begin{align}
	\tilde{\sigma}_{x} &= \mathcal{M}\sigma_x\mathcal{M}^{-1} = \sigma_z, \, 
	\tilde{\sigma}_{y} = \mathcal{M}\sigma_y\mathcal{M}^{-1} = (\bm{\sigma}\times\bm{\hat{n}})_z, \notag\\
	\tilde{\sigma}_{z} &= \mathcal{M}\sigma_z\mathcal{M}^{-1} = -\bm{\sigma}\cdot\bm{\hat{n}}, \,
		\tilde{\sigma}_{0} = \mathcal{M}\sigma_0\mathcal{M}^{-1} = \sigma_0. 
\end{align}

Similar generalization can be applied to the set of 16 $\Gamma$ matrices (see, for example Ref.~\onlinecite{PESKIN:1995Westview_QFT}) in the four-band description of bilayer graphene. The unitary transformation that can diagonalize the noninteracting Hamiltonian for bilayer graphene in Eq.~\eqref{Eq:BilayerH0} is given by
\begin{align}
 \mathcal{M} = 
 \begin{pmatrix}
 	\frac{\cos\theta}{\sqrt{2}} & \frac{\sin\theta}{\sqrt{2}} & -\frac{\sin\theta}{\sqrt{2}} & \frac{\cos\theta}{\sqrt{2}} \\
	e^{-i\phi}\frac{\sin\theta}{\sqrt{2}} & e^{-i\phi}\frac{\cos\theta}{\sqrt{2}} & 
	e^{-i\phi}\frac{\cos\theta}{\sqrt{2}} & -e^{-i\phi}\frac{\sin\theta}{\sqrt{2}} \\
	-\frac{\cos\theta}{\sqrt{2}} & \frac{\sin\theta}{\sqrt{2}} & \frac{\sin\theta}{\sqrt{2}} & \frac{\cos\theta}{\sqrt{2}} \\
	-e^{i\phi}\frac{\sin\theta}{\sqrt{2}} & e^{i\phi}\frac{\cos\theta}{\sqrt{2}} & 
	-e^{i\phi}\frac{\cos\theta}{\sqrt{2}} & -e^{i\phi}\frac{\sin\theta}{\sqrt{2}} \\
 \end{pmatrix},\label{Eq:UnitaryTransformation}
\end{align}
where $\theta\equiv\frac{1}{2}\tan^{-1}\left(\frac{2vk}{\gamma_1}\right)$, and $\phi=\tan^{-1}(\frac{k_y}{k_x})$. The above unitary matrix satisfies
\begin{align}
 \mathcal{M}^\dagger \mathcal{H}_0 \mathcal{M} = 
 \begin{pmatrix}
  \eps{1} & 0 & 0 & 0\\
  0 & \eps{2} & 0 & 0\\
  0 & 0 & \eps{3} & 0\\
  0 & 0 & 0 & \eps{4}
 \end{pmatrix}.
\end{align}
The four energy bands $\varepsilon_i(\bk)$ are given in Eq.~\eqref{Eq:BLband}. 

In addition, We will frequently use a set of modified $\Gamma$ matrices in our calculation. If the standard set of 16 $\Gamma$ matrices (see, for example Ref.~\onlinecite{PESKIN:1995Westview_QFT}) are denoted by $\Gamma_i$, then the ones we employ are 
\begin{align}
 \tilde{\Gamma}_i = \mathcal{M}\Gamma_i\mathcal{M}^\dagger,\quad i=1,2,\dots,16,
\end{align}
where the $\mathcal{M}$ is the unitary transformation introduced in Eq.~\eqref{Eq:UnitaryTransformation}. 
For future convenience, we will drop the tilde hereafter and it shall be understood that by $\Gamma$ matrices we always refer to this set of modified matrices. The density matrices encountered in our calculations will all be expanded in this transformed set of $\Gamma$ matrices.

\section{Thomas-Fermi screening wavevector in bilayer graphene\label{Appendix:Thomas-Fermi}}
The screening properties of an electron gas depend on the density of states $D_0$ at the Fermi level. If we use the simple Thomas-Fermi screening theory, the screening wavevector is given by 
\begin{align}
	\ktf=\dfrac{2\pi e^2}{\kappa}D_0=\dfrac{2\pi e^2}{\kappa} \partii{n}{\varf}, \label{Eq:DefOfKTF}
\end{align}
where $\kappa$ is the dielectric constant, $\varf$ is the Fermi energy, and $n$ is the electron density. In the four-band description, the electron density of bilayer graphene is 
\begin{align}
	n=\dfrac{\varf(\varf+\gamma_1)}{\pi\hbar^2v^2}+\Theta(\varf-\gamma_1)\dfrac{\varf(\varf-\gamma_1)}{\pi\hbar^2v^2},
\end{align}
where we have considered the four-fold spin-valley degeneracy. Therefore, the density of states $D_0$ is given by
\begin{align}
	D_0=\dfrac{(2\varf+\gamma_1)}{\pi\hbar^2v^2}+\Theta(\varf-\gamma_1)\dfrac{(2\varf-\gamma_1)}{\pi\hbar^2v^2}.
\end{align}
This result directly leads to the expression for $\ktf$ in Eq.~\eqref{Eq:TFScreening}. It is interesting to note that when the Fermi energy moves into the higher conduction band ($\varf\geq\gamma_1$), the electron density $n$ is continuous, whereas the density of states has a jump of $\gamma_1/\pi\hbar^2v^2$. Such a discontinuity is responsible for the jump at $\varf=\gamma_1$ in Fig.~\ref{Fig:DIC}. 

\section{The equilibrium self-energy\label{Appendix:SelfEnergy}}
It is instructive to write down explicitly the equilibrium self-energy in our model [see Eq.~\eqref{Eq:Self-Energy} for definitions]. In the band basis, the diagonal entries $\se_{\lambda}\equiv \bra{\lambda}\se(\bk)\ket{\lambda}$ represent the electron self-energy of each band. In addition, four of the off-diagonal entries are nonzero. The explicit expressions for these entries are presented below. 
\begin{widetext}
\allowdisplaybreaks[2]
\begin{align}
\se_{1}=& -\sum_{\bk'}V_{\bk\bk'} \Bigg\{\dfrac{1}{8}\left[(n_1+n_3)(3+\cos2\phi)+(n_2+n_4)(1-\cos2\phi)\right]+\dfrac{t}{8\delkp}(1-\cos2\phi)(n_1+n_2-n_3-n_4)\notag\\
 +\dfrac{t^2}{8\delk\delkp}&\left[(n_1-n_3)(3+\cos2\phi)-(n_2-n_4)(1-\cos2\phi)\right]+\dfrac{t}{8\delk}(1-\cos2\phi)(n_1-n_2+n_3-n_4)+\dfrac{2kk'\cos\phi}{\delk\delkp}(n_1-n_3)\Bigg\},\notag\\
\se_2=& -\sum_{\bk'}V_{\bk\bk'}
\Bigg\{\dfrac{1}{8}\left[(n_1+n_3)(1-\cos2\phi)+(n_2+n_4)(3+\cos2\phi)\right]+\dfrac{t}{8\delkp}(1-\cos2\phi)(n_1+n_2-n_3-n_4)\notag\\
 +\dfrac{t^2}{8\delk\delkp}&\left[(n_1-n_3)(\cos2\phi-1)+(n_2-n_4)(3+\cos2\phi)\right]+\dfrac{t}{8\delk}(1-\cos2\phi)(n_1-n_2+n_3-n_4)+\dfrac{2kk'\cos\phi}{\delk\delkp}(n_2-n_4)\Bigg\},\notag\\
\se_3 =& -\sum_{\bk'}V_{\bk\bk'}
\Bigg\{\dfrac{1}{8}\left[(n_1+n_3)(3+\cos2\phi)+(n_2+n_4)(1-\cos2\phi)\right]+\dfrac{t}{8\delkp}(1-\cos2\phi)(n_1+n_2-n_3-n_4)\notag\\
 +\dfrac{t^2}{8\delk\delkp}&\left[(n_3-n_1)(\cos2\phi+3)+(n_2-n_4)(1-\cos2\phi)\right]-\dfrac{t}{8\delk}(1-\cos2\phi)(n_1-n_2+n_3-n_4)-\dfrac{2kk'\cos\phi}{\delk\delkp}(n_1-n_3)\Bigg\},\notag\\
\se_4 =& -\sum_{\bk'}V_{\bk\bk'}
\Bigg\{\dfrac{1}{8}\left[(n_1+n_3)(1-\cos2\phi)+(n_2+n_4)(3+\cos2\phi)\right]-\dfrac{t}{8\delkp}(1-\cos2\phi)(n_1+n_2-n_3-n_4)\notag\\
 +\dfrac{t^2}{8\delk\delkp}&\left[(n_1-n_3)(1-\cos2\phi)-(n_2-n_4)(3+\cos2\phi)\right]-\dfrac{t}{8\delk}(1-\cos2\phi)(n_1-n_2+n_3-n_4)-\dfrac{2kk'\cos\phi}{\delk\delkp}(n_2-n_4)\Bigg\},\notag\\
\se_{13} \equiv& \bra{1}\se\ket{3} =-\sum_{\bk'}V_{\bk\bk'}\Bigg\{ \dfrac{vk\gamma_1}{4\delk\delkp}\left[(n_2-n_4)(1-\cos2\phi)-(n_1-n_3)(3+\cos2\phi)\right]
	+\dfrac{vk'\gamma_1}{\delk\delkp}(n_1-n_3)\cos\phi\notag\\&-\dfrac{vk}{4\delk}(n_1-n_2+n_3-n_4)(1-\cos2\phi)
	\Bigg\}=(\se_{31})^{\ast},\\
\se_{24} \equiv& \bra{2}\se\ket{4} =-\sum_{\bk'}V_{\bk\bk'}\Bigg\{ \dfrac{vk\gamma_1}{4\delk\delkp}\left[(n_2-n_4)(3+\cos2\phi)-(n_1-n_3)(1-\cos2\phi)\right]
-\dfrac{vk'\gamma_1}{\delk\delkp}(n_2-n_4)\cos\phi\notag\\&+\dfrac{vk}{4\delk}(n_1-n_2+n_3-n_4)(1-\cos2\phi)
	\Bigg\}=(\se_{42})^{\ast}.\notag
\end{align}
\end{widetext} 
Note that the Fermi distribution functions $n_i$ inside the integrals are all functions of $\bk'$. 

\section{Nonequilibrium self-energy changes \label{Appendix:EquationForG}}
One of the main results of this paper is the set of 16 equations in Eqs.~\eqref{Eq:FirstTwoEq_1}-\eqref{Eq:LastEq} in the main text. They completely determine the dynamics of the nonequilibrium density matrix $\rkl$ under an applied electric field. The right-hand-side of these equations are quite complicated and thus not given in the main text. In fact, they are non-equilibrium self-energy changes, which we list below:
\begin{widetext}
\begin{align}
	\dse_{23,-}&=[n_3(k)-n_2(k)]\sum_{\bk'}V_{\bk\bk'} \Big\{ \dfrac{B_{+}-B_{-}}{2\delk\delkp}(4v^2kk'\cos\phi+\gamma_1^2)+\dfrac{B_{+}+B_{-}}{2}+\dfrac{v\gamma_1(C_{+}-C_{-}) }{\delk\delkp} (k'-k\cos\phi)\notag\\&-\dfrac{vk(F_{+}+F_{-})}{\delk}\sin\phi\Big\},\notag\\
	\dse_{14,-}&=[n_1(k)-n_4(k)]\sum_{\bk'}V_{\bk\bk'}\Big\{\dfrac{B_{+}-B_{-}}{2\delk\delkp} (4v^2kk'\cos\phi+\gamma_1^2)-\dfrac{B_{+}+B_{-}}{2}+\dfrac{v\gamma_1(C_{+}-C_{-}) }{\delk\delkp} (k'-k\cos\phi)\notag\\&-\dfrac{vk(F_{+}+F_{-})}{\delk}\sin\phi\Big\},\notag\\
	\dse_{23,+}&=[n_3(k)-n_2(k)]\sum_{\bk'}V_{\bk\bk'} \Big\{ \dfrac{A_{+}+A_{-}}{4\delk\delkp}\left[\gamma_1(\gamma_1-\delk)+\gamma_1(\delk+\gamma_1)\cos2\phi+8v^2kk'\cos\phi\right]+\dfrac{H_{+}+H_{-}}{4\delk}(\delk+\gamma_1)\sin2\phi\notag\\ &-\dfrac{A_{+}-A_{-}}{4\delk}\left[(\gamma_1-\delk)-(\gamma_1+\delk)\cos2\phi\right]
	+\dfrac{D_{+}+D_{-}}{2\delk\delkp}v\left[k'(\gamma_1-\delk)+k'(\gamma_1+\delk)\cos2\phi-2k\gamma_1\cos\phi\right]\notag\\&+\dfrac{G_{+}-G_{-}}{4\delk\delkp}\left[\gamma_1(\gamma_1+\delk)\sin2\phi+8v^2kk'\sin\phi\right]+\dfrac{E_{+}-E_{-}}{2\delk\delkp}v\left[2k\gamma_1\sin\phi-k'(\gamma_1+\delk)\sin2\phi\right]
	\Big\},\notag\\
	\dse_{14,+}&=[n_4(k)-n_1(k)]\sum_{\bk'}V_{\bk\bk'} \Big\{ \dfrac{A_{+}+A_{-}}{4\delk\delkp}\left[\gamma_1(\gamma_1-\delk)+\gamma_1(\delk+\gamma_1)\cos2\phi+8v^2kk'\cos\phi\right]+\dfrac{H_{+}+H_{-}}{4\delk}(\delk+\gamma_1)\sin2\phi\notag\\ &-\dfrac{A_{+}-A_{-}}{4\delk}\left[(\gamma_1-\delk)-(\gamma_1+\delk)\cos2\phi\right]
			+\dfrac{D_{+}+D_{-}}{2\delk\delkp}v\left[k'(\gamma_1-\delk)+k'(\gamma_1+\delk)\cos2\phi-2k\gamma_1\cos\phi\right]\notag\\&+\dfrac{G_{+}-G_{-}}{4\delk\delkp}\left[\gamma_1(\gamma_1+\delk)\sin2\phi+8v^2kk'\sin\phi\right]+\dfrac{E_{+}-E_{-}}{2\delk\delkp}v\left[2k\gamma_1\sin\phi-k'(\gamma_1+\delk)\sin2\phi\right]
	\Big\},\notag\\
	\dse_{21,+}&=[n_1(k)-n_2(k)]\sum_{\bk'}V_{\bk\bk'} \Big\{
		\dfrac{A_{+}+A_{-}}{2\delk\delkp}\gamma_1v\left[k(1+\cos2\phi)-2k'\cos\phi\right]-\dfrac{A_{+}-A_{-}}{2\delk}vk(1-\cos2\phi)+\dfrac{D_{+}-D_{-}}{2}\cos\phi\notag\\&+\dfrac{D_{+}+D_{-}}{2\delk\delkp}\left[2v^2kk'(1+\cos2\phi)+\gamma_1^2\cos\phi\right]-\dfrac{E_{+}+E_{-}}{2\delkp}\gamma_1\sin\phi-\dfrac{E_{+}-E_{-}}{2\delk\delkp}(2v^2kk'\sin2\phi+\gamma_1^2\sin\phi)\notag\\&-\dfrac{G_{+}+G_{-}}{\delkp}vk'\sin\phi-\dfrac{G_{+}-G_{-}}{2\delk\delkp}\gamma_1v(2k'\sin\phi-k\sin2\phi)+\dfrac{H_{+}+H_{-}}{2\delk}vk\sin2\phi
	\Big\},\\
	\dse_{34,+}&=[n_3(k)-n_2(k)]\sum_{\bk'}V_{\bk\bk'} \Big\{
		\dfrac{A_{+}+A_{-}}{2\delk\delkp}\gamma_1v\left[k(1+\cos2\phi)-2k'\cos\phi\right]-\dfrac{A_{+}-A_{-}}{2\delk}vk(1-\cos2\phi)-\dfrac{D_{+}-D_{-}}{2}\cos\phi\notag\\&+\dfrac{D_{+}+D_{-}}{2\delk\delkp}\left[2v^2kk'(1+\cos2\phi)+\gamma_1^2\cos\phi\right]+\dfrac{E_{+}+E_{-}}{2\delkp}\gamma_1\sin\phi-\dfrac{E_{+}-E_{-}}{2\delk\delkp}(2v^2kk'\sin2\phi+\gamma_1^2\sin\phi)\notag\\&+\dfrac{G_{+}+G_{-}}{\delkp}vk'\sin\phi-\dfrac{G_{+}-G_{-}}{2\delk\delkp}\gamma_1v(2k'\sin\phi-k\sin2\phi)+\dfrac{H_{+}+H_{-}}{2\delk}vk\sin2\phi
	\Big\},\notag\\
		\dse_{21,-}&=[n_1(k)-n_2(k)]\sum_{\bk'}V_{\bk\bk'} \Big\{
		\dfrac{C_{+}+C_{-}}{2}\cos\phi+\dfrac{C_{+}-C_{-}}{2\delk\delkp}(4v^2kk'+\gamma_1^2\cos\phi)+\dfrac{F_{+}-F_{-}}{2}\sin\phi+\dfrac{F_{+}+F_{-}}{2\delk}\gamma_1\sin\phi\notag\\&+\dfrac{B_{+}-B_{-}}{\delk\delkp}\gamma_1v(k-k'\cos\phi)
	\Big\},\notag\\
	\dse_{34,-}&=[n_4(k)-n_3(k)]\sum_{\bk'}V_{\bk\bk'} \Big\{
		\dfrac{C_{+}+C_{-}}{2}\cos\phi-\dfrac{C_{+}-C_{-}}{2\delk\delkp}(4v^2kk'+\gamma_1^2\cos\phi)+\dfrac{F_{+}-F_{-}}{2}\sin\phi-\dfrac{F_{+}+F_{-}}{2\delk}\gamma_1\sin\phi\notag\\&-\dfrac{B_{+}-B_{-}}{\delk\delkp}\gamma_1v(k-k'\cos\phi)
		\Big\},\notag\\
	\dse_{13,-}&=[n_1(k)-n_3(k)]\sum_{\bk'}V_{\bk\bk'} \Big\{
		\dfrac{C_{+}+C_{-}}{2}\sin\phi+\dfrac{C_{+}-C_{-}}{2\delkp}\gamma_1\sin\phi-\dfrac{F_{+}-F_{-}}{2}\cos\phi-\dfrac{F_{+}+F_{-}}{2}\cos\phi\notag\\&-\dfrac{B_{+}-B_{-}}{\delkp}vk'\sin\phi
		\Big\},\notag\\
		\dse_{42,-}&=[n_2(k)-n_4(k)]\sum_{\bk'}V_{\bk\bk'} \Big\{
		\dfrac{C_{+}+C_{-}}{2}\sin\phi-\dfrac{C_{+}-C_{-}}{2\delkp}\gamma_1\sin\phi-\dfrac{F_{+}-F_{-}}{2}\cos\phi+\dfrac{F_{+}+F_{-}}{2}\cos\phi\notag\\&+\dfrac{B_{+}-B_{-}}{\delkp}vk'\sin\phi
		\Big\},\notag\\
	\dse_{13,+}&=[n_1(k)-n_3(k)]\sum_{\bk'}V_{\bk\bk'} \Big\{
		\dfrac{A_{+}+A_{-}}{2\delk\delkp}\gamma_1v(2k'\sin\phi-k\sin2\phi)-\dfrac{A_{+}-A_{-}}{2\delk}vk\sin2\phi-\dfrac{D_{+}-D_{-}}{2\delk}\gamma_1\sin\phi\notag\\&-\dfrac{D_{+}+D_{-}}{2\delk\delkp}(2v^2kk'\sin2\phi+\gamma_1^2\sin\phi)-\dfrac{E_{+}+E_{-}}{2\delk\delkp}(4v^2kk'+\gamma_1^2\cos\phi)-\dfrac{E_{+}-E_{-}}{2\delk\delkp}[2v^2kk'(1+\cos2\phi)+\gamma_1^2\cos\phi]\notag\\& -\dfrac{H_{+}+H_{-}}{2\delk}vk(1-\cos2\phi)+\dfrac{G_{+}+G_{-}}{\delk\delkp}\gamma_1v(k-k'\cos\phi)+\dfrac{G_{+}-G_{-}}{2\delk\delkp}\gamma_1v[k(1+\cos2\phi)-2k'\cos\phi]
		\Big\},\notag\\
		\dse_{42,+}&=[n_2(k)-n_4(k)]\sum_{\bk'}V_{\bk\bk'} \Big\{
		\dfrac{A_{+}+A_{-}}{2\delk\delkp}\gamma_1v(2k'\sin\phi-k\sin2\phi)-\dfrac{A_{+}-A_{-}}{2\delk}vk\sin2\phi+\dfrac{D_{+}-D_{-}}{2\delk}\gamma_1\sin\phi+\gamma_1^2\sin\phi)\notag\\&-\dfrac{D_{+}+D_{-}}{2\delk\delkp}(2v^2kk'\sin2\phi+\dfrac{E_{+}+E_{-}}{2\delk\delkp}(4v^2kk'+\gamma_1^2\cos\phi)-\dfrac{E_{+}-E_{-}}{2\delk\delkp}[2v^2kk'(1+\cos2\phi)+\gamma_1^2\cos\phi]\notag\\& -\dfrac{H_{+}+H_{-}}{2\delk}vk(1-\cos2\phi)-\dfrac{G_{+}+G_{-}}{\delk\delkp}\gamma_1v(k-k'\cos\phi)+\dfrac{G_{+}-G_{-}}{2\delk\delkp}\gamma_1v[k(1+\cos2\phi)-2k'\cos\phi]
		\Big\}.\notag
\end{align}
\end{widetext}
In the above expressions, the functions inside the integral, $A_{\pm}$, $B_{\pm}$, $C_{\pm}$, $D_{\pm}$, $E_{\pm}$, $F_{\pm}$, $G_{\pm}$, and $H_{\pm}$ are all functions of $\bk'$.

\section{Definition of some auxiliary functions\label{Appendix:AuxFunctions}}
The long-range Coulomb potential is given by 
\begin{align}
  V_{\bk\bk'} = \dfrac{2\pi e^2}{q_{\bk\bk'}}, \; q_{\bk\bk'} = \sqrt{\bk^2+\bk'^2-2kk'\cos\phi},
\end{align}
where $\phi$ is the angle formed by the two vectors $\bk$ and $\bk'$. When $k'\neq0$, the angular integral over $\phi$ is nontrivial, which have been discussed in more detail in Ref.~\onlinecite{Borghi:2009SSC_FermiVelocityEnhancement}. Here we just list the relevant ones below (assuming $k'\neq0$). We define $\Phi_n(k,k')$ to be the following integral, 
\begin{align}
	\Phi_n(k,k')\equiv  \int_0^{2\pi}d\phi\dfrac{\cos n\phi }{q_{\bk\bk'}},
\end{align}
with the understanding that $\cos(0\phi)\equiv 1$. 
We then have 
\allowdisplaybreaks[2]
 \begin{align}
\Phi_0(k,k') &=4\ellk(z)/(k+k'),\notag\\
 \Phi_1(k,k') &=\dfrac{2(k^2+k'^2)\ellk(z)-2(k+k')^2\elle(z)}{kk'(k+k')},\notag\\
  \Phi_2(k,k') &=4\Big[\dfrac{(k^4+k^2k'^2+k'^4)\ellk(z)}{3k^2k'^2(k+k')} \notag\\
 & \qquad-\dfrac{(k+k')^2(k^2+k'^2)\elle(z)}{3k^2k'^2(k+k')}\Big], \label{Eq:AngularIntegrals}\\
  \Phi_3(k,k')&=2\Big[\dfrac{(8k^6+7k^4k'^2+7k^2k'^4+8k'^6)\ellk(z)}{15k^3k'^3(k+k')}\notag\\
&\qquad-\dfrac{(k+k')^2(8k^4+7k^2k'^2+8k'^4)\elle(z)}{15k^3k'^3(k+k')}\Big]\notag,
\end{align}
where we have defined the dimensionless parameter $z=4kk'/(k+k')^2$. The $\ellk(z)$ ($\elle(z)$) is the complete elliptic integrals of the first (second) kind, defined as
\begin{align}
 \ellk(z) &= \int_0^{\pi/2} \dfrac{1}{\sqrt{1-z\sin^2\theta}}d\theta, \notag\\
 \elle(z) &= \int_0^{\pi/2}\sqrt{1-z\sin^2\theta}d\theta. \label{Eq:EllipticIntegrals}
\end{align}

\bibliography{CondBLG-References}
\end{document}